\documentclass[letterpaper,twocolumn,10pt]{article}
\pdfoutput=1
\usepackage{usenix}

\usepackage{times}

\usepackage{tikz}
\usepackage{amsmath}

\usepackage{amssymb}
\usepackage{pifont}%
\usepackage{tikz}

\usepackage{amsmath}
\usepackage{amsfonts}
\usepackage{amsthm}
\usepackage{subcaption}
\usepackage{multirow}
\usepackage{tabularx}
\usepackage{xcolor}
\usepackage[frozencache=true,cachedir=minted-cache]{minted}
\usepackage{xspace}
\usepackage{verbatim}
\usepackage[normalem]{ulem}
\usepackage{multicol}
\usepackage{float}
\usepackage{titlesec}
\usepackage{adjustbox}
\usepackage{authblk}
\usepackage{url}

\makeatletter
\renewcommand\AB@affilsepx{\hspace{0.5cm} \protect\Affilfont}
\makeatother

\usepackage[capitalize]{cleveref}
\crefformat{section}{#2\S#1#3}
\crefname{listing}{Lst.}{Lst.}

\newenvironment{tightitemize}{%
\begin{list}{$\bullet$}{%
\setlength{\itemsep}{1.5pt}%
\setlength{\topsep}{2pt}%
\setlength{\parskip}{0pt}%
\setlength{\parsep}{0pt}%

\setlength{\labelwidth}{0pt}%
\setlength{\leftmargin}{4pt}%
\setlength{\labelsep}{0pt}%
\setlength{\listparindent}{0pt}%
}}%
{\end{list}}

\usepackage{color}
\definecolor{codegreen}{rgb}{0,0.6,0}
\definecolor{codered}{rgb}{0.7,0.2,0}
\definecolor{codegray}{rgb}{0.5,0.5,0.5}
\definecolor{codepurple}{rgb}{0.58,0,0.82}
\definecolor{backcolour}{rgb}{0.95,0.95,0.92}
\definecolor{purple}{RGB}{128,0,128}
\definecolor{indigo}{RGB}{75,0,130}
\definecolor{royalblue}{RGB}{65,105,225}
\definecolor{navy}{RGB}{0,0,128}
\usepackage[ruled,vlined]{algorithm2e}

\newcommand*\justify{%
  \fontdimen2\font=0.4em%
  \fontdimen3\font=0.2em%
  \fontdimen4\font=0.1em%
  \fontdimen7\font=0.1em%
  \hyphenchar\font=`\-%
}

\renewcommand{\texttt}[1]{%
  \begingroup
  \ttfamily
  \begingroup\lccode`~=`/\lowercase{\endgroup\def~}{/\discretionary{}{}{}}%
  \begingroup\lccode`~=`[\lowercase{\endgroup\def~}{[\discretionary{}{}{}}%
  \begingroup\lccode`~=`.\lowercase{\endgroup\def~}{.\discretionary{}{}{}}%
  \catcode`/=\active\catcode`[=\active\catcode`.=\active
  \justify\scantokens{#1\noexpand}%
  \endgroup
} 

\newtheorem{theorem}{Theorem}[section]
\newtheorem{lemma}[theorem]{Lemma}

\newif\ifcommenton
\commentonfalse
\ifcommenton
\newcommand{\alexey}[1]{\textcolor{royalblue}{\textbf{AT: #1}}} %
\newcommand{\alind}[1]{\textcolor{codegreen}{\textbf{AK: #1}}}
\newcommand{\dhruv}[1]{\textcolor{purple}{\textbf{DG: #1}}}
\newcommand{\snigdha}[1]{\textcolor{navy}{\textbf{SG: #1}}}
\newcommand{\brian}[1]{}
\newcommand{\sukrit}[1]{\textcolor{magenta}{\textbf{Suk: #1}}}
\else
\newcommand{\alexey}[1]{}
\newcommand{\alind}[1]{}
\newcommand{\dhruv}[1]{}
\newcommand{\brian}[1]{}
\newcommand{\snigdha}[1]{}
\newcommand{\sukrit}[1]{}

\fi

\newcommand{\system}{SuperServe\xspace}
\newcommand{\sysname}{SuperServe\xspace}
\newcommand{\sysmech}{SubNetAct\xspace}
\newcommand{\syspol}{SlackFit\xspace}

\newcommand{\ie}{{i.e.,}~}

\newcommand{\wrt}{{w.r.t.}~}

\newcommand{\infaas}{{InFaaS}~}

\newcommand{\secref}[1]{\S\ref{#1}}
\newcommand{\figref}[1]{Fig.~\ref{#1}}

\newcommand{\myparagraph}[1]{\noindent{\bfseries #1.}}

\newcommand{\one}{({\em i}\/)}
\newcommand{\two}{({\em ii}\/)}
\newcommand{\three}{({\em iii}\/)}

\begin{document}

\date{}

\title{
\sysname{}: Fine-Grained Inference Serving for Unpredictable Workloads
}

\author[1]{Alind Khare}
\author[1]{Dhruv Garg}
\author[2]{Sukrit Kalra}
\author[3]{Snigdha Grandhi$^{*}$\thanks{Work done as a student at Georgia Tech}}
\author[2]{Ion Stoica}
\author[1]{Alexey Tumanov}
\affil[1]{Georgia Tech}
\affil[2]{UC Berkeley}
\affil[3]{Adobe}
\affil[ ]{\authorcr \{alindkhare, dgarg39, sgrandhi32, atumanov\}@gatech.edu}
\affil[ ]{\authorcr \{sukrit.kalra, istoica\}@berkeley.edu}

\maketitle
\begin{abstract}
The increasing deployment of ML models on the critical path of production applications in both
datacenter and the edge requires ML inference serving systems to serve these models under
unpredictable and bursty request arrival rates.
Serving models under such conditions requires these systems to strike a careful
balance between the latency and accuracy requirements of the application and the overall
efficiency of utilization of scarce resources.
State-of-the-art systems resolve this tension by either choosing a static point
in the latency-accuracy tradeoff space to serve all requests or 
load specific models on the critical path of request serving.

In this work, we instead resolve this tension by simultaneously
serving the entire-range of models spanning the latency-accuracy tradeoff space.
Our novel mechanism, \sysmech{}, achieves this by carefully inserting specialized
operators in weight-shared SuperNetworks.
These operators enable \sysmech{} to dynamically route requests
through the network to meet a latency and accuracy target.
\sysmech{} requires upto $2.6\times$ lower memory to serve a vastly-higher number of
models than prior state-of-the-art.
In addition, \sysmech{}'s near-instantaneous actuation of models unlocks
the design space of fine-grained, reactive scheduling policies.
We explore the design of one such extremely effective policy, \syspol{} and instantiate
both \sysmech{} and \syspol{} in a real system, \sysname{}.
\sysname achieves $4.67$\% higher accuracy for the same SLO attainment and $2.85\times$ higher SLO attainment for the same accuracy on a trace derived from the real-world Microsoft Azure Functions workload and yields the best tradeoffs on a wide range of extremely-bursty synthetic traces \textit{automatically}.

\end{abstract}

\section{Introduction}

Recent advancements in machine-learning (ML) techniques have unlocked vast improvements in both accuracy and efficiency of
a variety of tasks such as image classification~\cite{image_classification_1}, object detection~\cite{image_segmentation, object_recognition}, text summarization~\cite{text_summarization}, sentiment analysis~\cite{sentiment_analysis}, next word prediction~\cite{next-word} etc.
As a result, ML models have been quickly deployed across a wide-range of applications in both datacenters~\cite{zhang2019accelerating, soifer2019deep, fb-datacenter, applied-fb, tfserving} and on the edge~\cite{ananthanarayanan2017real, padmanabhan2023gemel, self_driving, lin2018architectural}, and are now subject to the stringent requirements of a production application.

Notably, ML models on the critical-path of these applications must now deal with \emph{unpredictable}
request rates that rapidly change at a sub-second granularity.
For example, web applications in datacenters increasingly rely on ML models~\cite{zhang2019accelerating, soifer2019deep}, 
and are known to have extremely bursty request rates, with a peak demand that is $50\times$ higher than the average~\cite{li2023alpaserve}.
Similarly, request rates in autonomous vehicles change rapidly as a function of various factors such as the
terrain (city vs. freeway driving), time of the day etc.~\cite{self_driving}.

Thus, ML inference serving systems have the arduous task of striking a careful balance between three key requirements of production applications facing unpredictable request rates:

\myparagraph{R1: Latency} 
ML models are being increasingly deployed in applications that have extremely stringent latency requirements, 
quantified using a Service-Level Objective (SLO)~\cite{jyothi2016morpheus}.
For example, both web serving~\cite{soifer2019deep, fb-datacenter, clockwork} in datacenters and 
autonomous vehicles~\cite{self_driving, lin2018architectural} on the edge must maximize the number of requests completed 
within the specified SLO ranging from 10s to 100s of milliseconds~\cite{self_driving, zhang2023shepherd}.

\myparagraph{R2: Accuracy} 
Techniques such as Neural Architecure Search (NAS)~\cite{mnasnet, nasRL, baker2016designing} have enabled the development of multiple
ML models that offer varying accuracies for a particular task.
As a result, applications demand the highest-accuracy results possible
within the latency targets of their requests.
For example, increased accuracy has been intricately tied to a better user experience for web applications~\cite{wu2019machine, applied-fb}.
Similarly, the safety of an autonomous vehicle heavily relies on the
accuracy of various different ML models~\cite{self_driving, gog2021pylot}.

\myparagraph{R3: Resource-Efficiency} 
Web applications at Facebook process $200$ trillion ML model 
requests daily~\cite{venkatapuram2020custom}, which
represents a significant fraction of Facebook's datacenter 
demands~\cite{fb-datacenter}.
In addition to the increasing proliferation of ML models, 
their growing  reliance on scarce resources such as GPUs 
and specialized accelerators (e.g, TPUs~\cite{google_tpu}, AWS 
Inferentia~\cite{aws_inferentia}) for efficient inference has lead to
resource tensions across applications in both datacenters and on the 
edge~\cite{padmanabhan2023gemel}.
Thus, inference serving systems that serve a 
wide range of ML models must make judicial use of these scarce resources.

The first-generation of inference serving systems~\cite{clipper, tfserving, inferline, sagemaker, triton, clockwork, zhang2023shepherd}
resolves this tension by choosing a static point in the tradeoff-space between \textbf{R1}-\textbf{R3}
and serving all requests using the same model for the entirety of the application's runtime.
As a result, applications must make a one-time decision to forego meeting their SLO targets (\textbf{R1}) under bursty 
request rates or suffer degraded accuracy (\textbf{R2}) under normal conditions.
More recently, state-of-the-art inference serving systems~\cite{infaas, modelswitch} enable applications to register
multiple ML models spanning the entire pareto frontier of latency (\textbf{R1}) and accuracy (\textbf{R2}) targets, and 
\emph{automatically} choose the appropriate model to serve requests based on the incoming request rates.
These systems 
must either keep the entire set of models in memory or rely on \emph{model switching} techniques to load
the required models at runtime~\cite{modelswitch}.
As GPU memory remains the key resource bottleneck in both datacenter and edge inference 
serving~\cite{li2023alpaserve, padmanabhan2023gemel}, these systems must now choose between 
\textbf{R3} -- effectively utilizing the available resources (by incurring the enormous latency penalties of 
switching models), or
\textbf{R1} -- meeting SLO targets under highly unpredictable request rates.

Conventional wisdom in inference serving literature touts the ``\emph{non-negligible provisioning time} [for ML models due to switching], 
\emph{which can exceed the request processing times}" as a 
``\emph{key characteristic of ML workloads}", and "\emph{rules out reactive techniques}"
for responding to bursty request rates~\cite{swayam}.
This wisdom has been widely accepted~\cite{infaas, clipper, clockwork} leading to the development of coarse-grained
scheduling policies for inference serving that must account for the enormous latency penalty of switching models when 
reacting to bursty request rates.
As a result, these coarse-grained policies typically avoid or minimize switching models by design~\cite{infaas}, and
are hence, unable to optimally navigate the tradeoff space between \textbf{R1}-\textbf{R3}
under rapidly-changing, unpredictable request rates.

In this work, we challenge this conventional wisdom that forces
a choice between \textbf{R1} and \textbf{R3}.
We describe a mechanism, \sysmech{},
to simultaneously serve the entire range of models
spanning the latency-accuracy tradeoff space (\textbf{R1}-\textbf{R2})
in a resource-efficient manner (\textbf{R3}).
At the core of our mechanism are novel control-flow and slicing operators
that \sysmech{} carefully inserts into the SuperNet~\cite{ofa, compofa} neural architectures.
SuperNets enable a latency-accuracy tradeoff
(\textbf{R1}-\textbf{R2}) by training  a set of shared model weights 
for many neural networks,  without duplication.
Prior works~\cite{ofa, compofa} propose efficient mechanisms for
training SuperNets for both vision and NLP tasks, but require 
each model instance to be individually extracted for inference, 
leading to a similar choice as before between \textbf{R1} and \textbf{R3} -- either load all individual models or switch between them at runtime.
However, \sysmech{}'s novel operators obviate the need to
extract individual models and load them dynamically at runtime.
Instead, \sysmech{} dynamically routes requests within one 
SuperNet deployment with negligible overhead, enabling 
near-instantaneous \emph{actuation} of different models.
As a result, \sysmech{} unlocks orders of magnitude improvements
in the navigation of the latency-accuracy tradeoff-space 
(\textbf{R1}-\textbf{R2}), while substantially  reducing the memory 
footprint (\textbf{R3})  (see \cref{sec:motivation}).

In addition to being resource-efficient (\textbf{R3}), \sysmech{}'s agility in navigating 
the latency-accuracy tradeoff space (\textbf{R1}-\textbf{R2}) fundamentally 
changes the design space of scheduling policies.
Instead of complex scheduling policies that must reason
about future request rates in a bid to avoid paying the latency
of switching ML models dynamically under bursts, \sysmech{} enables
the specification of simple policies that 
directly optimize for the key success metrics 
\textbf{R1}-\textbf{R3}.
While conventional wisdom deems such reactive policies infeasible, 
we explore one example point in this design space with a simple, yet effective
policy that we call \syspol{}.
\syspol{} is a reactive scheduling policy that exploits
the near-instantaneous actuation property of \sysmech{} to
make fine-grained decisions about how many requests to
serve in a batch, and which latency/accuracy choice to
select for serving in real-time.

We summarize the contributions of this paper as follows:
\begin{tightitemize}
    \item We introduce \sysmech{} (\cref{sec:sysarch}), a novel mechanism that enables a resource-efficient, 
    fine-grained navigation of the latency-accuracy tradeoff space. \sysmech{} achieves this by carefully
    inserting novel control-flow and slicing operators that dynamically route requests  through
    a single SuperNet.
    \item We unlock the design space of fine-grained, reactive scheduling policies and provide a mathematical
    formulation of their objective (\cref{sec:policy:problem}). We then propose \syspol{} (\cref{sec:pol:slackfit}), a simple, yet effective greedy heuristic
    and show how it accurately approximates the optimal objective.
    \item We instantiate \sysmech{} and \syspol{} in a real-world system, \sysname{}, a real-time asynchronous model serving system with pluggable scheduling policies (\cref{sec:sysarch:arch}).
    \item We extensively evalute \sysname{} with both \syspol{} and several state-of-the-art scheduling policies
    (\cref{sec:eval}). We find that \sysname{} achieves $4.67$\% higher accuracy for the same SLO attainment and $2.85\times$ higher SLO attainment for the same accuracy on the real-world Microsoft Azure Functions trace. %
\end{tightitemize}

\label{sec:intro}

\section{Motivation and Background}
\begin{figure*}[t!]
	\centering
	\begin{subfigure}[b]{0.293\textwidth}
        \includegraphics[width=\textwidth]{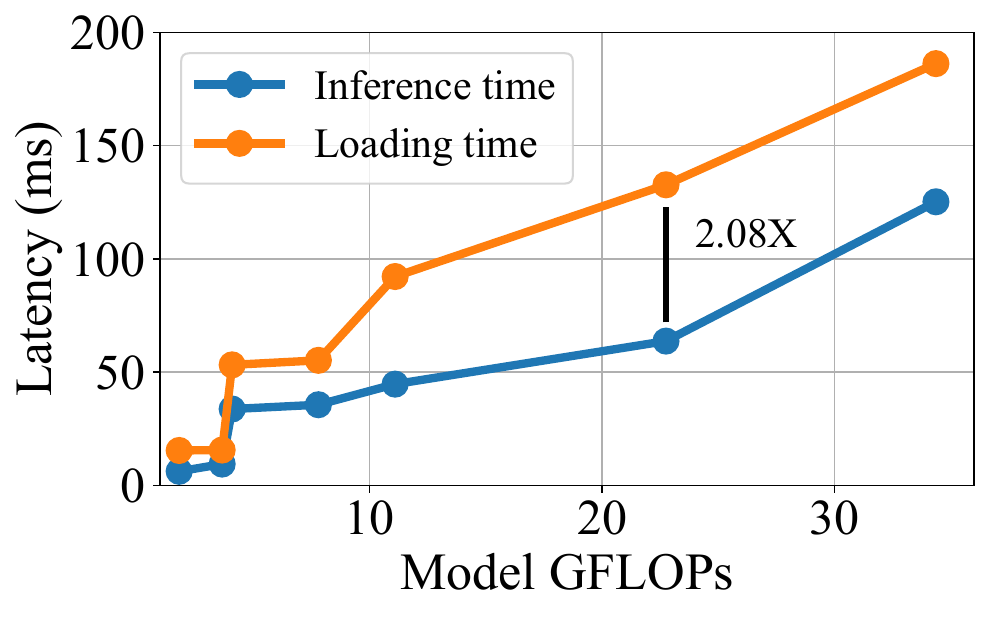}
	    \caption{
		     Model switching is expensive
		}
		\label{fig:motiv:model_infer_vs_load}
	\end{subfigure}
	\begin{subfigure}[b]{0.3\textwidth}
        \includegraphics[width=\textwidth]{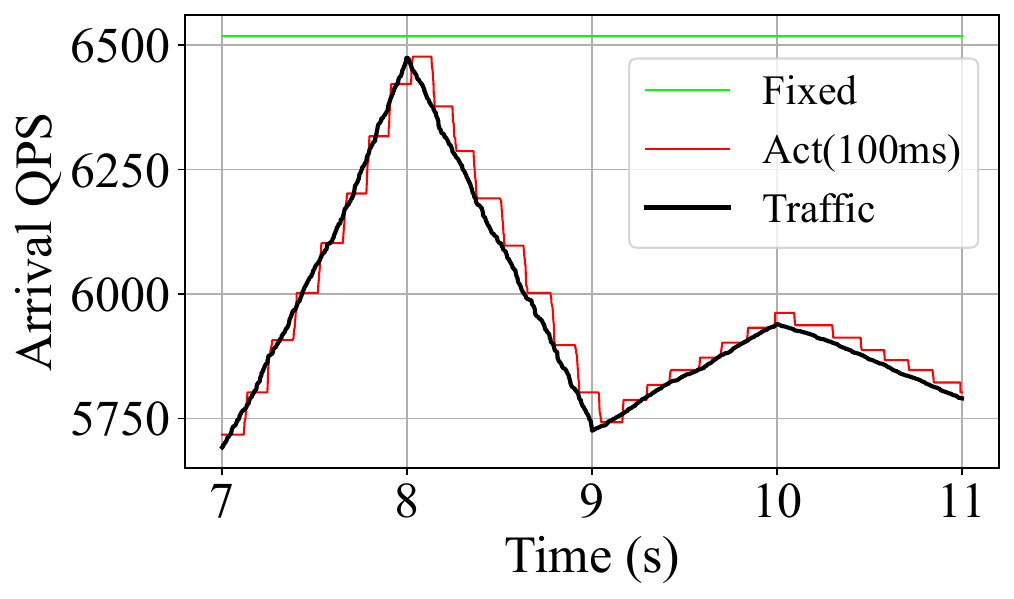}
	    \caption{
		    Coarse-grained scheduling
		}
		\label{fig:motiv:act_delay_100ms}
	\end{subfigure}
	\begin{subfigure}[b]{0.3\textwidth}
        \includegraphics[width=\textwidth]{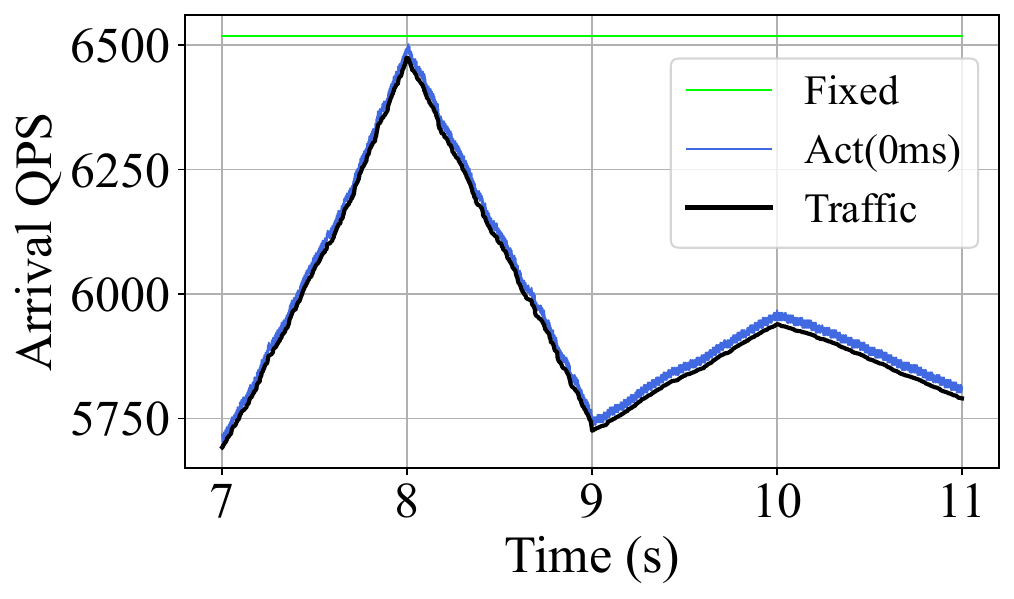}
	    \caption{
		  Fine-grained scheduling
		}
		\label{fig:motiv:act_delay_0ms}
	\end{subfigure}
    \caption{\small 
        \textbf{Fine-grained scheduling policies are beneficial.}  
        (a) The latency of loading ResNet-18,34,50,101,152 \protect\cite{resnet}, Wide-Resnet-101 \protect\cite{wide_resnet} and ConvNeXt-Large \protect\cite{convnext}
        is greater than its inference latency with a batch size of 16, making model
        switching expensive. This gap widens as model sizes increase.
	    (b) A bursty request rate from the MAF~\protect\cite{maf} trace with a coarse-grained scheduling policy leads to 2\% of the requests missing their SLO (\textbf{R1}) 
        and wasting resources (\textbf{R3}) due to a $100$ms actuation delay typical of model switching techniques.
    	(c) A bursty request rate with a fine grained scheduling policy that can switch models instantaneously can achieve 0\% SLO misses and effectively utilizes the GPU.
    }
	 \label{fig:motiv:benefit:model_switching}
  \vspace{-0.2in}
\end{figure*}

In \cref{sec:motivation:bursty}, we motivate the development of a reactive, fine-grained scheduling policy
that maximizes
\textbf{R1}-\textbf{R3} under
unpredictable, bursty request rates.
\cref{sec:motivation:supernetworks} then provides a background on Supernets~\cite{ofa, compofa} and discusses the properties of Supernets 
relevant to \textbf{R1}-\textbf{R2}
that make them a good fit for a fine-grained exploration of the latency-accuracy tradeoff.

\subsection{Fine-Grained Reactive Scheduling}
\label{sec:motivation:bursty}
Prior works in inference serving systems~\cite{zhang2023shepherd, li2023alpaserve} have exhaustively
analyzed both production traces from Microsoft Azure Functions (MAF)~\cite{maf} and synthetic 
application traces with a goal of highlighting their bursty request arrival patterns.
For example, Zhang et al.~\cite{zhang2023shepherd} 
underscore the high coefficient of variance in request arrivals in production
traces~\cite{maf}.
Further, the authors claim that the bursty ``\emph{sub-second request arrival
patterns} [are] \emph{nearly impossible to predict}", thus frustrating the goal
of meeting the stringent SLO requirements of requests in an ML-based production applications.

A strawman solution to fulfilling SLOs under bursty request rates requires
inference serving systems to provision for the peak.
In this setting, these systems load the entire set of models spanning the
latency-accuracy tradeoff space into GPU memory
and switch between them as request rate fluctuates.
While this reduces the actuation latency of switching models, allowing inference
serving systems to rapidly degrade accuracy (\textbf{R2}) under bursts to meet SLO
targets (\textbf{R1}), it wastes resources under normal request rates (\textbf{R3}). 

As a result, state-of-the-art inference serving systems~\cite{modelswitch, infaas}
rely on model switching techniques that page models in and out when
required, to make efficient use of GPU memory (\textbf{R3}).
However, as we show in \cref{fig:motiv:model_infer_vs_load}, the loading
time of ML models from CPU to GPU memory is vastly more than the inference time of
the biggest batch size, and the gap widens as the model sizes increase.
Thus, in a bid to offset the cost of loading models on the critical path of
handling requests, inference serving systems rely on \emph{predictive} scheduling policies 
that make coarse-grained estimations of future request arrival patterns.
Such policies are bound to be suboptimal owing to the difficulty of predicting
the short bursts in request arrival rates coupled with their stringent SLO requirements~\cite{zhang2023shepherd}.

We believe that the key to optimally serving bursty request rates instead
lies in the ability to rapidly switch between ML models thus obviating the
need for coarse-grained predictive scheduling policies.
To validate our hypothesis, we simulate a coarse-grained policy with
an actuation delay (i.e., time taken to switch to a new ML model that can handle
the current request rate) of $100$ms and an idealistic fine-grained  policy
with an actuation delay of $0$ms.
\cref{fig:motiv:act_delay_100ms,fig:motiv:act_delay_0ms} plot the effects of these
policies on a small bursty subtrace from the MAF trace.
We observe that the coarse-grained policy leads to higher SLO misses (\textbf{R1}) 
under increasing request rates and wasted resources (\textbf{R3}) under decreasing
request rates.
On the other hand, the fine-grained policy is able to instantaneously adjust to
the increasing and decreasing request rates leading to no missed SLOs and effective
utilization of the GPU.

\begin{figure}[t!]
	\centering
    \includegraphics[width=0.7\columnwidth]{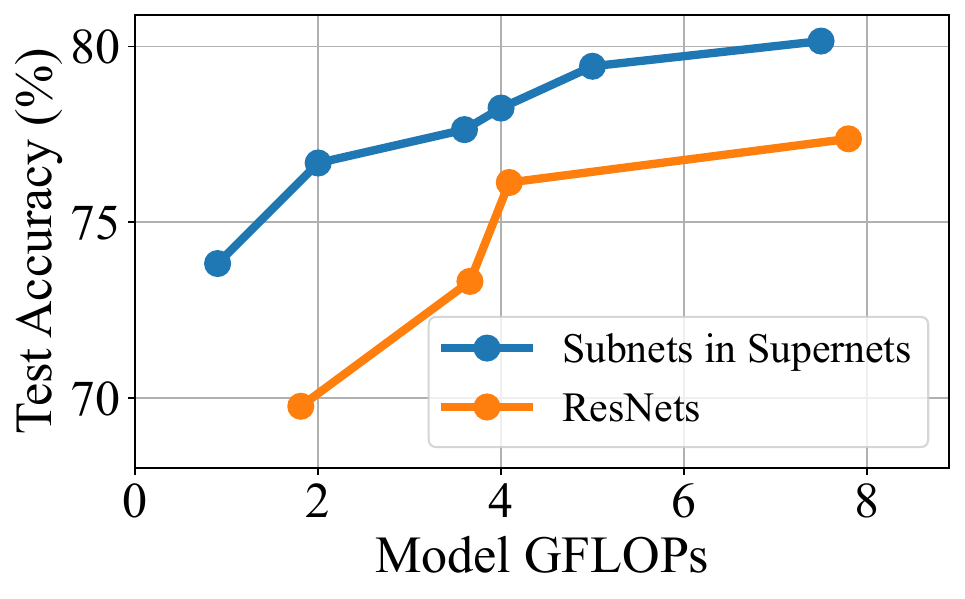}
    \caption{\small 
        \textbf{Supernets provide better latency-accuracy tradeoff.} 
        The accuracy (\textbf{R2}) of subnets individually extracted from OFAResNet supernet~\protect\cite{ofa} is vastly superior to the hand-tuned ResNets from \cref{fig:motiv:model_infer_vs_load} for the same FLOP requirements (\textbf{R1}).
        }
	 \label{fig:motiv:benefit:lat_acc_tradeoff}
	 \vspace{-0.2in}
\end{figure}

\subsection{Weight-Shared Supernets}
\label{sec:motivation:supernetworks}
The problem of navigating the pareto-optimal frontier of the latency-accuracy
tradeoff space (\textbf{R1}-\textbf{R2}) by finding highest accuracy DNNs for a specific latency target is well studied in ML literature.
Conventional Neural Architecture Search (NAS)~\cite{nasnet, nasRL, mnasnet, proxylessnas, darts} approaches have enabled architectures tailored to a particular latency target.
However, these approaches search for and train individual networks for a particular latency
target and hence require inference serving systems to either load all of them
into memory and waste resources (\textbf{R3}), or switch between them at runtime.

Instead, recent works~\cite{ofa, compofa} have proposed first training one \emph{supernet}
and then extracting a subset of its layers to form \emph{subnets}.
As a result,
once the supernet is trained, no further retraining is required for a
specific subnet.
Each extracted subnet targets a specific point in the latency-accuracy tradeoff 
space, and partially shares its weights/layers with other subnets.
In addition, this automated neural architecture search (NAS)
yields subnets that correspond to vastly superior points in the latency-accuracy tradeoff space (\textbf{R1}-\textbf{R2}).
For example, \cref{fig:motiv:benefit:lat_acc_tradeoff} highlights the accuracy benefits
of subnets extracted from a ResNet-based supernet when
compared to the hand-tuned ResNets for an equivalent number of FLOPs. 

In order to find subnets that target a particular point in the latency-accuracy tradeoff space, the architecture search in supernets relies on the following parameters: 
- \one{} \textit{Depth} ($\mathbb{D}$) describes the depth of a subnet, 
   \two{} \textit{Expand Ratio} ($\mathbb{E}$) describes layer-wise ratio of output to input channels of a convolution or fully-connected layer, and 
   \three{} \textit{Width Multiplier} ($\mathbb{W}$) describes layer-wise fraction of input and output channels to be used. 
These parameters ($\mathbb{D}$, $\mathbb{E}$, $\mathbb{W}$) combinatorially create an architecture space, $\Phi$ ($|\Phi| \approx 10^{19}$)~\cite{ofa}, from which individual subnets are extracted \emph{statically} for inference.

\label{sec:back}
\label{sec:motivation}
\label{sec:motiv}

\section{\sysmech{}: Instantaneous Model Actuation}
Motivated by \cref{sec:motivation:bursty}, we seek to develop a fine-grained, reactive scheduling mechanism
that enables efficient execution of ML-based production applications (\textbf{R1}-\textbf{R3}) under bursty request rates.
As discussed in \cref{sec:motivation:supernetworks}, prior work in supernets~\cite{ofa, compofa} enables the extraction of individual models that target a specific point in the latency-accuracy
tradeoff space (\textbf{R1}-\textbf{R2}).
However, these approaches yield individual models that must either be simultaneously deployed 
(wasting resources; \textbf{R3}) or paged in as request rates fluctuate (missing SLOs; \textbf{R1}).

To resolve this fundamental tension, we make the key observation that by virtue of performing architectural
search \emph{post} training, a supernet subsumes the entire architectural space of subnets.
As a result, instead of extracting and deploying individual subnets (as done previously), we can instead
deploy a supernet and \emph{dynamically route} requests to the appropriate subnet.
This observation leads us to introduce \sysmech{}, a memory-efficient model \emph{actuation}
mechanism (\textbf{R3}) that exposes fine-grained control decisions to near-instantaneously switch between 
subnets in order to pick the optimal point in the latency-accuracy tradeoff space (\textbf{R1}-\textbf{R2}).

\begin{figure}[t!]
	\centering
    \includegraphics[width=\columnwidth]{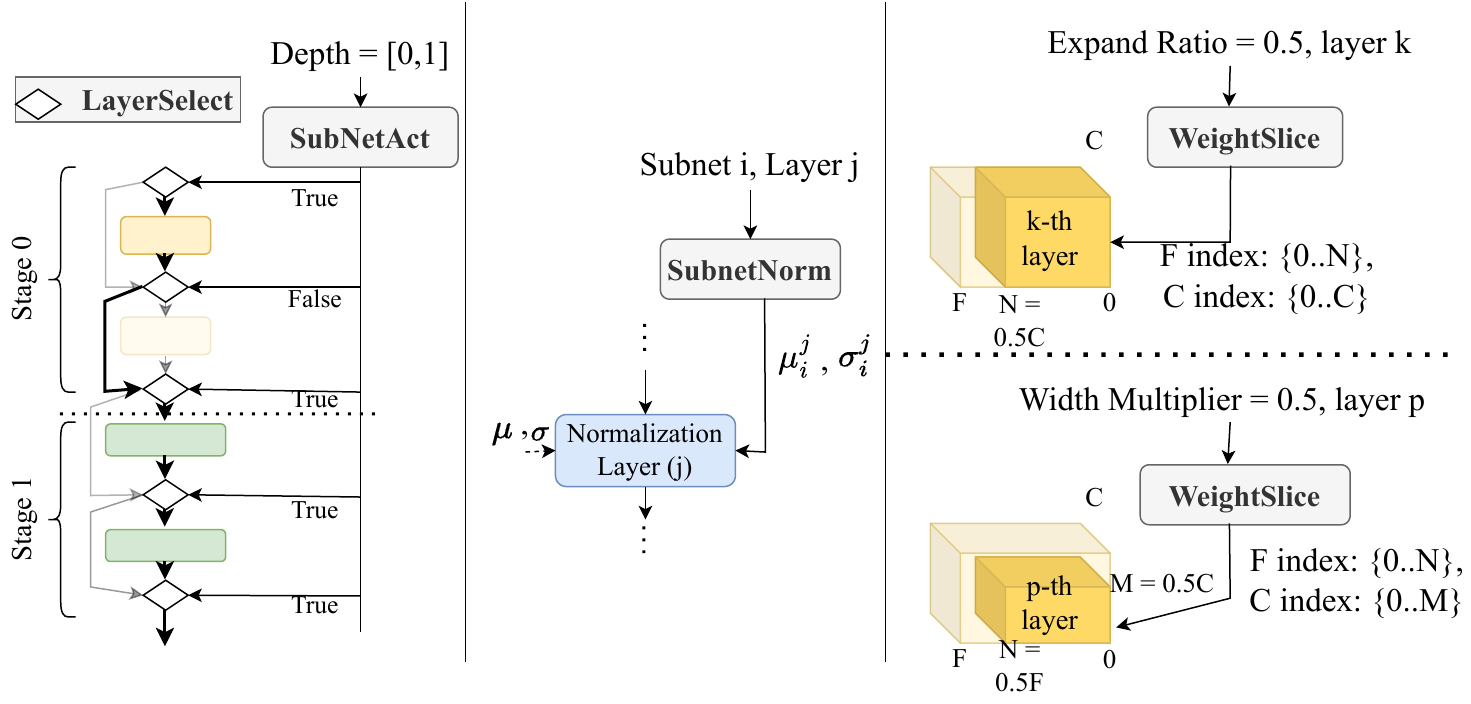}
    \caption{\small 
        \textbf{\sysmech's novel operators} (LayerSelect, SubnetNorm, WeightSlice) dynamically actuate subnets by routing requests through weight-shared layers and non-weight-shared components.
    }
	 \label{fig:sys_mech}
	 \vspace{-0.2in}
\end{figure}
\begin{figure}[tb!]
	\centering
    \includegraphics[width=\columnwidth,height=1.9cm, keepaspectratio]{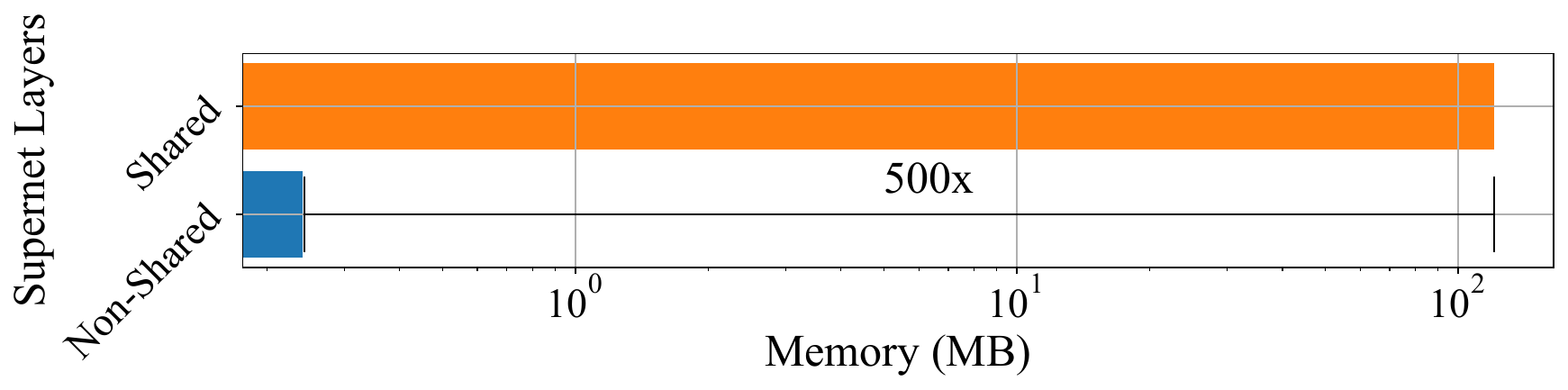}
    \caption{\small 
        \textbf{\sysmech's memory savings.}  
        The memory used by the normalization statistics is 
        $500\times$ smaller than the non-normalization layers.
        SubnetNorm decouples the normalization statistics for
        each subnet and provides accurate bookeeping thus enabling high accuracy (\textbf{R1}), with minimal increase in memory consumption (\textbf{R3}).
    }
	 \label{fig:subnet_act:mem_consumption}
	 \vspace{-0.2in}
\end{figure}

\sysmech{}'s (\cref{fig:sys_mech}) key insight lies in the introduction of the following three novel operators that enable it to dynamically route requests to the required subnet in a supernet:

\begin{figure*}[t!]
	\centering
	\begin{subfigure}[b]{0.3\textwidth}
        \includegraphics[width=\textwidth]{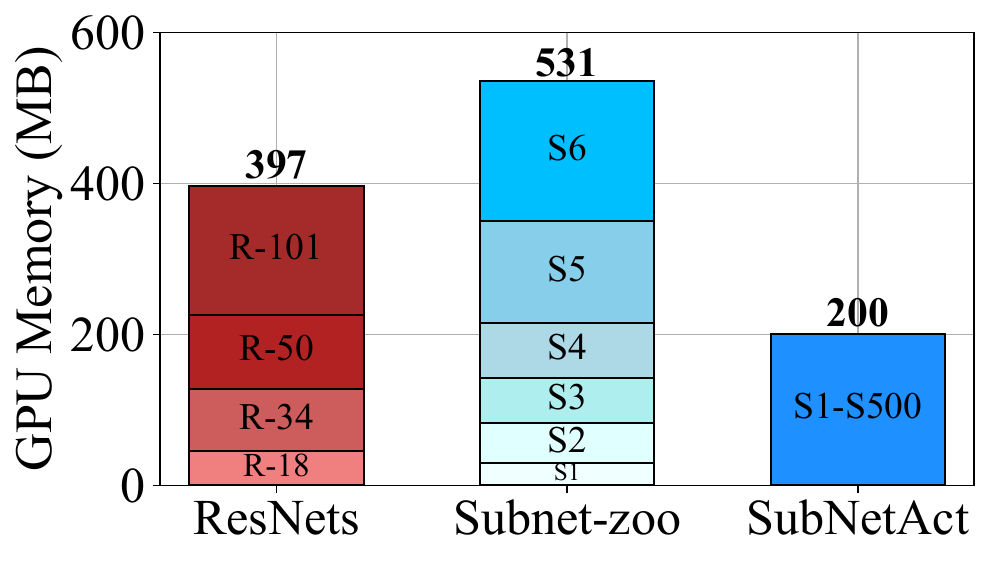}
	    \caption{
		     Reduced Memory Requirements.
		}
		\label{fig:subnetact:benefit:resource_efficiency}
	\end{subfigure}
	\begin{subfigure}[b]{0.3\textwidth}
        \includegraphics[width=\textwidth]{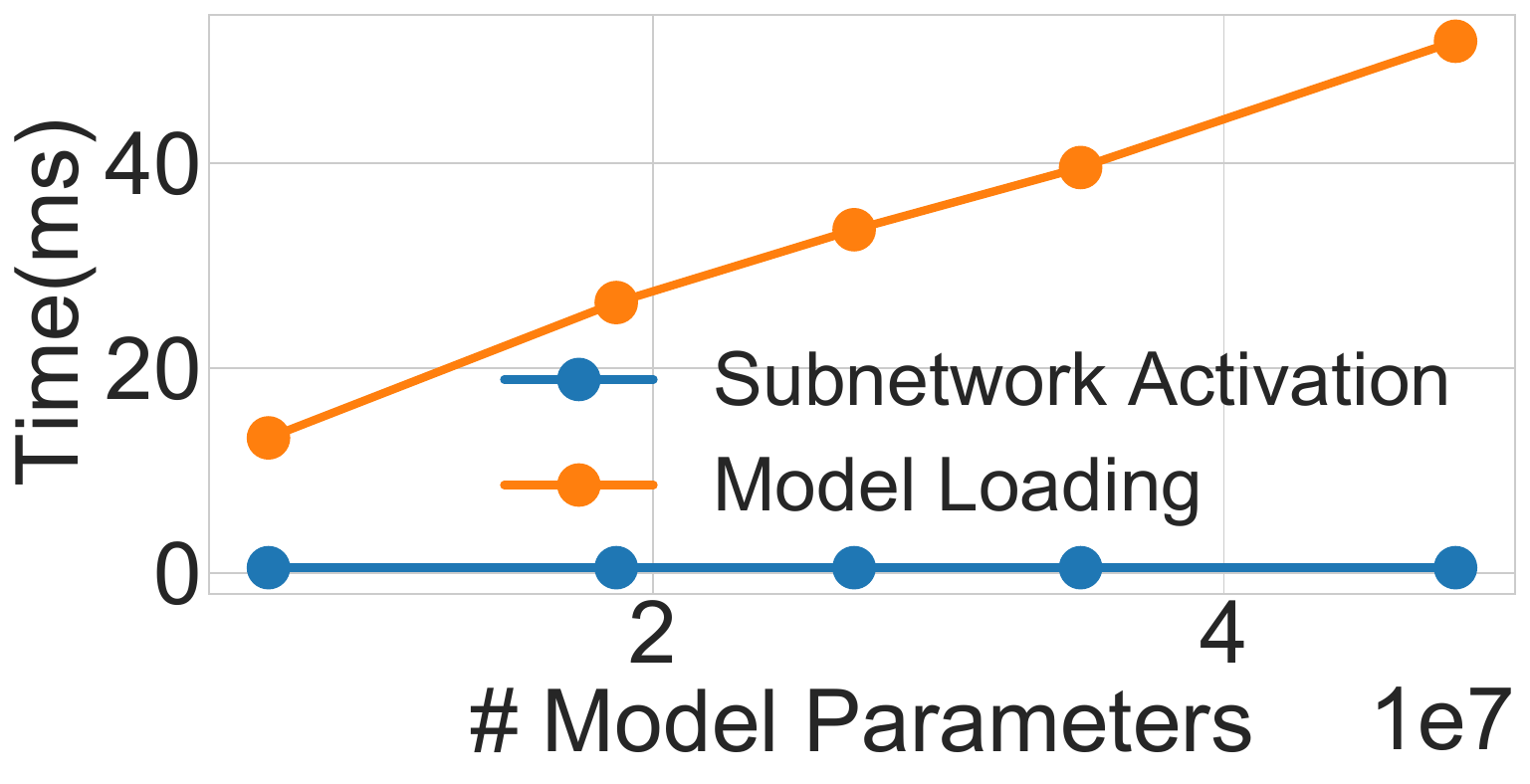}
	    \caption{
		    Instantaneous Model Actuation.
		}
		\label{fig:subnetact:benefit:instant_model_switch}
	\end{subfigure}
	\begin{subfigure}[b]{0.3\textwidth}
        \includegraphics[width=\textwidth]{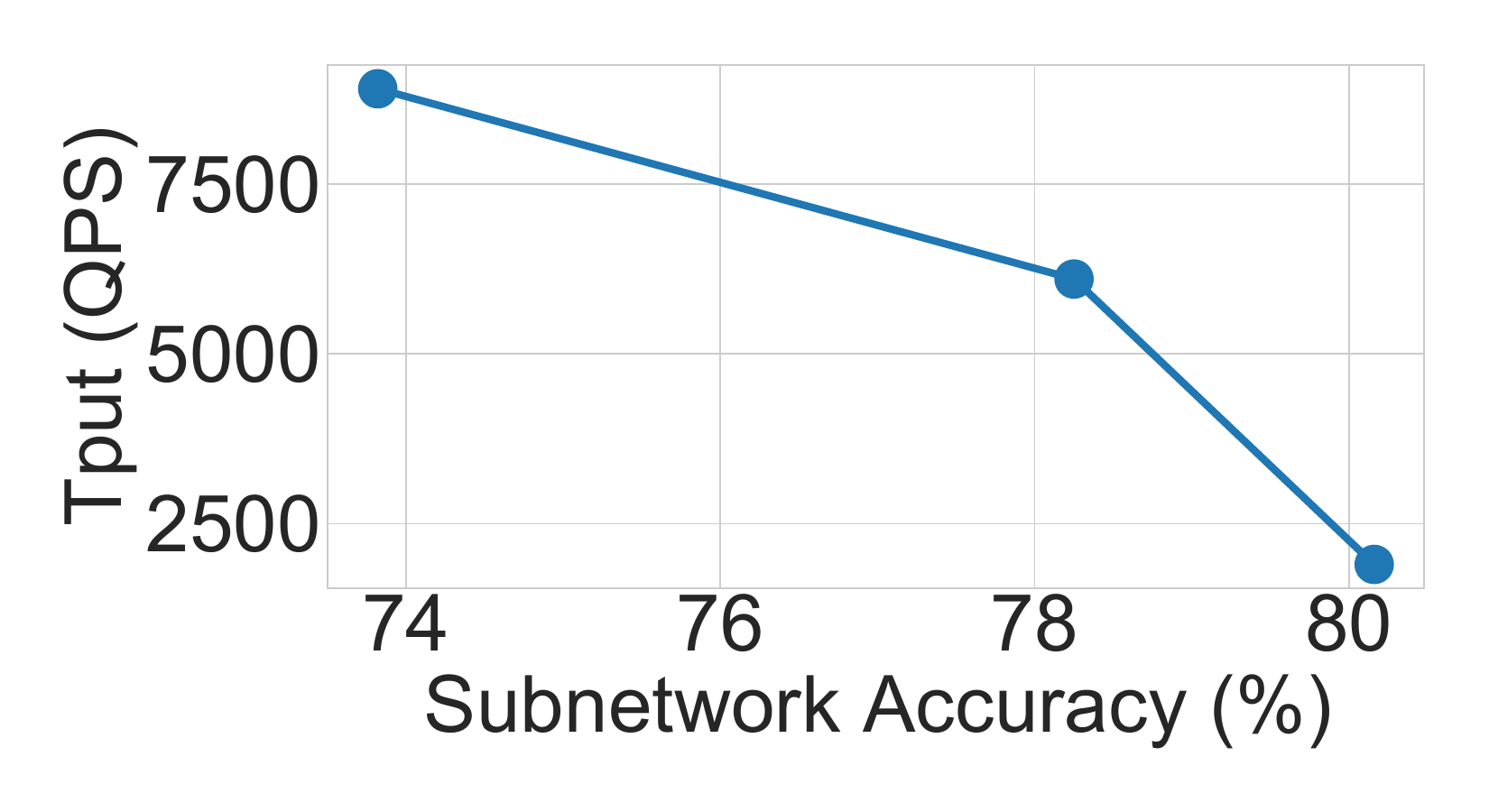}
	    \caption{
		  High dynamic throughput range. %
		}
		\label{fig:subnetact:benefit:th_range}
	\end{subfigure}
    \caption{\small 
        \textbf{Efficacy of \sysmech.} %
        (a) \sysmech{} requires upto $2.6\times$ lower memory to serve a higher-range of models when
        compared to the ResNets from \cref{fig:motiv:model_infer_vs_load} and six individual subnets 
        extracted from supernet~\protect\cite{ofa}
	    (b) \sysmech{} actuates different subnets near-instantaneously ($<1$ms), which is orders of
        magnitude faster than the model switching time.
    	(c) \sysmech{} enables instantaneous actuation of models that can sustain higher ingest rates thus inducing a wide dynamic throughput range ($\approx 2-8k$ queries per second) and increased accuracy. 
    }
	 \label{fig:motiv:benefit}
  \vspace{-0.2in}
\end{figure*}

\textbf{LayerSelect.} \sysmech{} takes as input the depth $\mathbb{D}$, and dynamically executes layers of a specific subnet based on $\mathbb{D}$. 
The depth $\mathbb{D}$ is converted by \sysmech{} to per-layer boolean values that determine which layers participate in inference.
\sysmech{} then wraps each layer in a LayerSelect operator that enforces control-flow by either 
passing the input activation to the wrapped layer or skipping the layer and directly forwarding input to the next layer.
This operator enables layer-sharing among subnets that differ in depth ($\Phi_{\mathbb{D}} \subset \Phi$), which reduces
the GPU memory consumption of the supernet (\textbf{R3}).
Moreover, it enables near-instantaneous (\textbf{R1}) switching of the supernet's accuracy (\textbf{R2}) under bursty request rates.

\textbf{SubnetNorm.} 
We observe that naively introducing the LayerSelect operator leads to a significant drop in subnet accuracy
(as low as $10$\%).
This is due to the incorrect tracking of the mean ($\mu$) and variance ($\sigma$) in normalization
layers such as BatchNorm~\cite{batchnorm}.
To account for this discrepancy, \sysmech{} introduces the SubnetNorm operator that precomputes and stores
$\mu$ and $\sigma$ for each possible subnet by performing forward pass inference on the training data.
SubnetNorm takes as input a unique subnet ID ($i$) and a layer ID ($j$) and outputs the precomputed normalization statistics $\mu_{i,j}$ and $\sigma_{i,j}$.
The layer $j$ then uses the provided statistics to perform normalization of activations,
effectively specializing $j$ for each subnet $i$.

Although this bookkeeping increases the memory requirements
of deploying the supernet, \cref{fig:subnet_act:mem_consumption} shows that the
overhead of these non-shared normalization statistics is $500\times$ smaller than
the memory requirement of the shared layers.
As a result, \sysmech{} can host thousands of subnets in memory by only keeping
the statistics unique to each subnet and sharing the non-normalization
weights amongst all the subnets.

\textbf{WeightSlice.}  This operator dynamically selects channels in convolution or fully-connected layers during inference. The input to WeightSlice is the expand ratio ($\mathbb{E}$) or width multiplier ($\mathbb{W}$)
for each layer, which collectively determine the number of channels to be used.
The operator outputs layer-specific channel indices, which are then used to select weight subsets for the forward pass inference.
The operator enables partial layer-sharing among subnets ($\{\Phi_{\mathbb{E}} \cup \Phi_{\mathbb{W}} \} \subset \Phi$), thus increasing the number of available subnet architectures.
As a result, \sysmech{} is able to provide the entire set of 
latency-accuracy options (\textbf{R1}-\textbf{R2}) to the scheduling policies.

We note that the input to these operators (i.e., depth ($\mathbb{D}$), expand ratio ($\mathbb{E}$) and width multiplier ($\mathbb{W}$)) remains similar to the inputs for 
architectural search in ML literature~\cite{ofa}.
Moreover, these control inputs are independent from the input to the actuated subnet (i.e., the request served by the
model), and are declaratively specified by a scheduling policy 
(\cref{sec:fine-grained-scheduler}).
Given the arrival rate, the scheduling policy chooses a specific subnet for a request (by specifying the control tuple
$\mathbb{D}$, $\mathbb{E}$ and $\mathbb{W}$), which is then actuated by \sysmech{} near-instantaneously.

\subsection{Discussion: Efficacy of \sysmech{}}
\label{sec:mech:micro}
We now highlight \sysmech{}'s efficacy in achieving key application requirements (\textbf{R1}-\textbf{R3})
under bursty request rates.

\myparagraph{Reduced Memory Requirements}
\sysmech{}'s novel operators enable subnets to share layers in place and dynamically route requests to the
appropriate subnet based on the control tuple $\mathbb{D}$, $\mathbb{E}$ and $\mathbb{W}$ determined by a scheduling policy.
As a result, \sysmech{} can simultaneously serve the entire range of models spanning the latency-accuracy
tradeoff space while drastically reducing memory requirements.
\cref{fig:subnetact:benefit:resource_efficiency} demonstrates this by comparing the memory requirement
of loading four different ResNets~\cite{resnet}, six individually extracted subnets~\cite{ofa}, and \sysmech that enables simultaneous actuation of $500$ subnets.
We observe that \sysmech{} can reduce memory consumption by upto $2.6\times$, while vastly increasing
the  latency-accuracy tradeoff points that can be actuated.

\myparagraph{Near-Instantaneous Model Actuation}
While switching between individual models requires loading their weights to the GPU, \sysmech{}'s operators
enable scheduling policies to actuate any subnet \textit{in place} without incurring additional loading overhead.
\cref{fig:subnetact:benefit:instant_model_switch} compares the time taken to perform on-demand loading of individual subnetworks versus in-place actuation of a subnet in \sysmech.
We observe that \sysmech{}'s model actuation is orders-of-magnitude faster than on-demand loading of ML models.
This allows scheduling policies that use \sysmech{} to rapidly actuate lower-accuracy models under
bursty conditions (\textbf{R1}) and switch to higher-accuracy models under normal load (\textbf{R2}), 
without coarse-grained predictions about future request rates.

\myparagraph{Increased Throughput \& Accuracy}
By providing instant model actuation, \sysmech{} allows scheduling policies to scale the throughput of the system up and down rapidly, thus inducing a broad throughput range within as few as 6\% of accuracy to help meet SLO guarantees (\textbf{R1}-\textbf{R2}).
\cref{fig:subnetact:benefit:th_range} compares the maximum sustained ingest throughput for a point-based open-loop arrival curve for serving the largest, smallest, and a median subnetwork on 8 \sysmech workers{}.  
We observe that \sysmech{} can serve a wide throughput range from $2000$-$8000$ QPS, while being able to instantaneously increase
accuracy between $74\%$ to $80\%$.
\label{sec:sysarch}
\section{Fine-Grained Scheduling Policies}
\sysmech{}'s near-instantaneous actuation of the entire latency-accuracy tradeoff space unlocks
the development of extremely fine-grained scheduling policies that can directly optimize for
\textbf{R1}-\textbf{R2}.
We note that by virtue of aggressively sharing weights/layers and dynamically routing requests
within a supernet, \sysmech automatically maximizes resource efficiency (\textbf{R3}).
In this section, we first start with the problem formulation of online serving for the  
new space of fine-grained policies enabled by \sysmech (\cref{sec:policy:problem}).
We then describe our proposed policy \syspol that aims to achieve both high accuracy 
and latency SLO attainment (\cref{sec:pol:slackfit}).

\subsection{Problem Formulation}
\label{sec:policy:problem}
\sysmech{} hosts a set of all possible subnets $\Phi$, where each subnet $\phi \in \Phi$ 
is defined by the control tuple $\mathbb{D}$, $\mathbb{E}$, $\mathbb{W}$ and has an accuracy $Acc(\phi)$ and a latency-profile $l_{\phi}(B)$ as a function of batch-size $B$.
A \emph{query}\footnote{We use the term \emph{query} and \emph{request} interchangeably.} $q$ arrives
to the scheduler at time $a_q$ with an SLO $d_q$.
A fine-grained scheduling policy focuses on selecting a subnet across supernets for queries across GPUs,
along with a batch $B$ to execute the query in~\cite{clipper, clockwork}.
The arrival time $a(B)$ of $B$ is the earliest arrival time and the deadline $d(B)$ is the earliest deadline of all queries in $B$. 
$\mathcal{B}$ is the set of all possible batches of queries. 
The policy decides if $B \in \mathcal{B}$ should execute at time $t$ using subnet $\phi$ on GPU $n$, which
is captured by a decision variable $I(B,t,n,\phi) \in \{0,1\}$. 

\myparagraph{Goal} The policy's goal is to maximize the number of accurate responses (to queries) within specified SLO (\textbf{R1}-\textbf{R2}).

\myparagraph{Optimal Offline ILP} 
We now present the ILP formulation of a scheduling policy that achieves our stated goal with 
an oracular knowledge about future query arrivals:
\begin{subequations}\label{eq:ilp}
\begin{equation}
    \text{maximize} \sum_t \sum_n \sum_{\phi \in \Phi}\sum_{B \in \mathbb{B}} \text{Acc}(\phi)\cdot|B| \cdot I(B,t,n,\phi) \tag{\ref{eq:ilp}}
\end{equation}
\begin{equation}
\text{s.t.}  \sum_t \sum_n \sum_{\phi \in \Phi} \sum_{\{B | q \in \mathbb{B}\}}  I(B,t,n,\phi) \leq 1, \;\;\;\;\;\;\;  \forall q 
\end{equation}
\begin{equation}
\sum_{B \in \mathbb{B}}    \sum_{ \{t^{'}\leq t \leq t^{'}+l_{\phi}(B) \}}      I(B,t^{'},n,\phi) \leq 1, \;\;\;\;\;\;\;  \forall n,t,\phi 
\end{equation}
\begin{equation}
    a(B)\cdot I(B,t,n,\phi) \leq t, \;\;\;\;\;\;\;  \forall n,t,B,\phi
\end{equation}
\begin{equation}
    \sum_{\phi \in \Phi} I(B,t,n,\phi) \leq 1, \;\;\;\;\;\;\;  \forall n,t,B
\end{equation}
\begin{equation}
    \sum_{\phi \in \Phi} (l_{\phi}(B)+t) \cdot I(B,t,n,\phi) \leq d(B), \;\;\;\;\;\;\;  \forall n,t,B
\end{equation}
\begin{equation}
    I(B,t,n,\phi) \in \{ 0, 1\}, \;\;\;\;\;\;\; \forall n,t,B,\phi
\end{equation}
\end{subequations}
The ILP maximizes the number of queries that satisfy their latency SLOs with the highest possible accuracy across all the selected query batches \ie $\exists \, \phi : I(B,t,n,\phi) = 1$ and $Acc(\phi) \cdot |B|$ is maximized. The constraints of the ILP denote -
\begin{tightitemize}
    \item[(1a)] A query $q$ can be assigned to at-most one batch $B$.
    \item[(1b)] A GPU $n$ can only execute a single subnet $\phi$ on a single batch $B$ at a particular time $t$.
    \item[(1c)] Batch $B$ can only execute after its arrival time $a(B)$.%
    \item[(1d)] Each batch $B$ can be served with a maximum of one subnet $\phi$ on a GPU $n$ at a time $t$. 
    \item[(1e)] The batch should complete before deadline $d(B)$.
    \item[(1f)] The choice variable $I(B,t,n,\phi)$ is a boolean indicator.
\end{tightitemize}
We note that our formulation \eqref{eq:ilp} is a Zero-one Integer Linear Program (ZILP) and solving
it is known to be NP-Hard~\cite{zilp, zhang2023shepherd}.
Furthermore, it is impractical to expect oracular query arrival knowledge. This renders the use of the formulated ILP in the online model serving setting unrealistic.
Instead, we approximate its behavior
under different query traffic conditions with an \textit{online} scheduling policy instead.
\begin{figure}[t!]
	\centering
	\begin{subfigure}[b]{0.3\textwidth}
        \includegraphics[width=\textwidth]{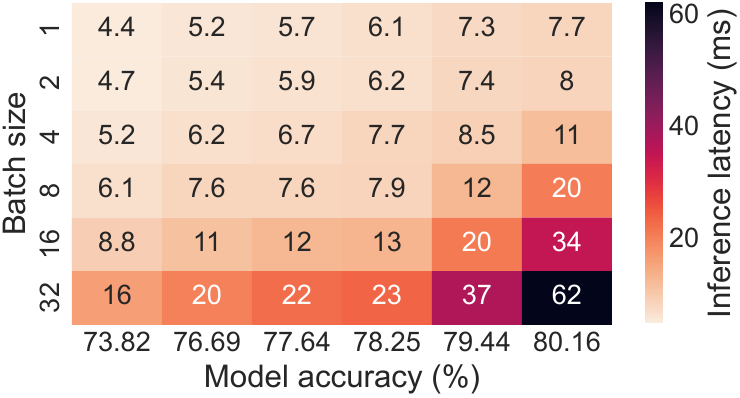}
	    \caption{
		     Latency Heatmap
		}
	\label{fig:lat_profile:heatmap}
	\end{subfigure}
	\begin{subfigure}[b]{0.15\textwidth}
        \includegraphics[width=\textwidth]{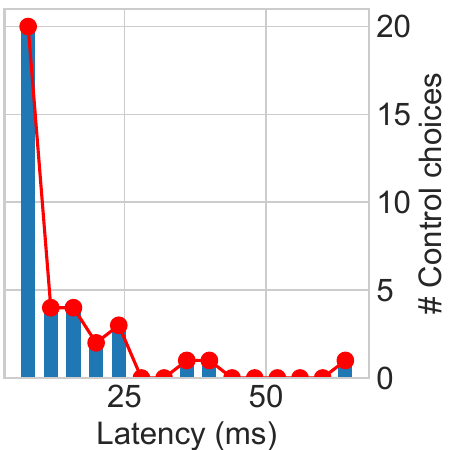}
	    \caption{
		    Control Choices 
		}
	\label{fig:lat_profile:num_choices}
	\end{subfigure}
    \caption{\small
        \textbf{\syspol control parameter space.} 
        (a) latency heatmap for six different (uniformly sampled \wrt FLOPs) pareto-optimal subnets in \sysmech as a function of accuracy (x-axis) and batch size (y-axis). 
        The latency is \textit{monotonic} with batch size and accuracy. 
        (b) the number of available choices
        decrease as latency increases.
    }
	 \label{fig:lat_profile:all}
	 \vspace{-0.2in}
\end{figure}

\subsection{\syspol: Online-Scheduling Policy}
\label{sec:pol:slackfit}

We introduce \syspol---a simple yet effective online scheduling policy that aims to maximize accuracy and latency SLO (\textbf{R1}-\textbf{R2}). \syspol is a greedy heuristic that approximates the ILP-based policy in Eq. \eqref{eq:ilp} and makes the decision-making tractable. \syspol makes following design choices:

\myparagraph{Operates on Pareto-Optimal Subnets ($\Phi_{\text{pareto}}$)} To make subnet choices in reasonable time, \syspol operates on the $\Phi_{\text{pareto}}$ instead of $\Phi$. $\Phi_{\text{pareto}}$ is a set of pareto-optimal subnets \wrt latency, accuracy obtained by using existing neural architecture search (NAS) methods\cite{ofa}\footnote{It takes $\leq$ 2min to perform NAS on supernets by using latency and accuracy predictors}. The size of $|\Phi_{\text{pareto}}| \approx 10^3$ is orders of magnitude smaller than $|\Phi| \approx 10^{19}$. This contributes to rapid scheduling decisions in \syspol.

\myparagraph{Uses Monotonic Properties of Subnets in $\Phi_{\text{pareto}}$} \syspol leverages key properties of subnets in $\Phi_{\text{pareto}}$ to further reduce the control search space by performing $O(log)$ operations. \figref{fig:lat_profile:heatmap} shows latency profiles of six subnets  uniformly sampled \wrt FLOPs from  $\Phi_{\text{pareto}}$. The properties of subnets in $\Phi_{\text{pareto}}$ are - (\textbf{P1}) the  latency increases monotonically with batch size,
(\textbf{P2}) the  latency increases monotonically with accuracy,  (\textbf{P3}) the latency difference among different batch sizes of a subnet increases with increase in subnet accuracy.

Properties \textbf{P1} and \textbf{P2} further reduce the dimensionality of control search for \syspol to a single dimension (\ie latency) to determine both the subnet $\phi$ and batch size $B$. Instead of searching the two-dimensional table (\figref{fig:lat_profile:heatmap}), 
\syspol operates on one-dimensional set of evenly sized latency buckets (\figref{fig:lat_profile:num_choices}). 
Each bucket consists of control tuples $(B,\phi)$ such that $l_{\phi}(B)$ remains within the range of bucket width. 
Buckets are constructed to exploit \textbf{P3} and to further reduce the search complexity to $O(1)$ for bucket selection. By construction and property \textbf{P3},
low latency buckets contain lower accuracy, higher throughput control choices, as the inference latency of smaller accuracy subnets on higher batch sizes is relatively lower. High latency buckets contain higher accuracy, lower throughput control choices.

\myparagraph{Slack-Based Decision-making}   \syspol's insight is that the remaining slack of the most urgent query provides proxy to changes in the traffic. Traffic peaks lead to more queueing delays in the system and in-turn reduces the remaining slack. Under low traffic, the slack remains high. Therefore, \syspol uses remaining slack for decision making.
It chooses a bucket with latency that is closest to and less than the remaining slack of the most urgent query. All control choices within the bucket satisfy query deadline. It picks the control choice with maximum batch size to opt for a high throughput choice especially for lower slack (traffic peaks).

\myparagraph{\syspol Behavior}
With slack-based decision making, \syspol can automatically adjust accuracy  (\textbf{R2}) and throughput of the system by choosing appropriate latency bucket on variable arrival traffic to maintain high SLO (\textbf{R1}). 
A well-behaved trace (e.g., low ingest rate, variation) results in higher slack. Higher slack leads to the choice of higher latency buckets. And, higher latency buckets  are correlated strongly with the probability of choosing higher accuracy models (due to property \textbf{P2}). 
Conversely, bursty traces lead to lower latency bucket choices, as \syspol operates under reduced latency slack. Lower latency bucket choices have higher batch sizes configuration (due to property \textbf{P3}). Thus, \syspol opportunistically maximizes accuracy while satisfying latency SLO.

\subsubsection{\syspol Approximation of Optimal Offline ILP}
\label{sec:pol:approx}
We now provide insights on how \syspol emulates behavior of the optimal offline ILP. To understand the behavior of ILP, we formulate a proxy utility function that captures the inner-term of the ILP objective function in Eq \ref{eq:ilp}, the utility function is defined for all subnet $\phi$, batch-size $B$ and a deadline $d_B$ (earliest deadline of queries in batch) --
\begin{equation}
    \mathbb{U}(\phi,B,d_B) = \begin{cases}
\text{Acc}(\phi) \cdot |B|, &\text{if } l_{\phi}(B) < d_B\\
0, &\text{otherwise}
\end{cases}
\label{eq:slackutil}
\end{equation}
The utility in Eq.~\ref{eq:slackutil} is non-zero if and only if subnet $\phi$ performs batch-inference on batch $B$ within the deadline $d_B$, and is zero otherwise. Note that this captures the success metrics: maximizing both the number of queries processed (\textbf{R1}) within their deadline and the accuracy they were served with (\textbf{R2}).

\noindent \textbf{A. Offline ILP and \syspol Prefer Pareto-Optimal Subnets.}
\syspol's key design choice is to operate pareto-optimal subnets \wrt latency, accuracy ($\Phi_{\text{pareto}}$) (\secref{sec:pol:slackfit}). We claim that offline ILP also tends to pareto-optimal subnets (\wrt latency, accuracy), as the these subnets yield higher utility.
\begin{lemma}
    The utility of pareto-optimal subnets is higher than non-pareto-optimal subnets if they have similar inference latency for a batch of queries.
    \begin{equation*}
    \begin{split}
        \mathbb{U}(\phi_p,B,d_B) > \mathbb{U}(\phi_q,B,d_B), \;\;\;\;\;\;\;\;\forall B,d_B \\ 
        \text{s.t.}\;\; \phi_p \in \Phi_{\text{pareto}},\; \phi_q \in \{\Phi \setminus \Phi_{\text{pareto}}\}, \; l_{\phi_p}(B) \approx l_{\phi_q}(B) 
    \end{split}
    \end{equation*}
\end{lemma}
This validates \syspol's design choice to operate on pareto-optimal subnets only. We defer the proof to Appendix \secref{sec:pareto:utility}. 

\noindent \textbf{B. Offline ILP and \syspol Prioritize Lower Accuracy \& Higher Batch Size under High Load.}
   We make a key observation that the utility of lower accuracy $\text{Acc}(\phi_{low})$, higher batch sizes ($B_{high}$) configurations is higher than higher accuracy ($\text{Acc}(\phi_{high})$), lower batch size ($B_{low}$) configuration in pareto-subnets of \sysmech.  This is because the factor difference in accuracy of pareto-subnets ($< 1$) is less than the factor differences of batch-sizes as seen in \figref{fig:lat_profile:heatmap} \ie $\frac{\text{Acc}(\phi_{high})}{\text{Acc}(\phi_{low})} \leq \frac{|B_{high}|}{|B_{low}|} \Rightarrow \text{Acc}(\phi_{high}) \cdot |B_{low}| \leq \text{Acc}(\phi_{low}) \cdot |B_{high}| $. Therefore, $\mathbb{U}(\phi_{low},B_{high}, d_q) \geq \mathbb{U}(\phi_{high},B_{low},d_q)$ may hold true under high load, in cases where the most urgent query q in a batch of $k$ queries ($q \in B_k$) can be served either by a) low accuracy model ($\phi_{min}$) with batch size $B_k$ or  b) higher accuracy model  ($\phi_{max}$)  on a subset of queries (say $m$, $q \in B_m$) with remaining queries ($B_k \setminus B_m$) missing the deadline due to high load. In such cases, the offline ILP will tend to option (a). \syspol also tends to lower accuracy and higher batch size options under heavy load, as described in ``\syspol's Behavior'' (\secref{sec:pol:slackfit}).

   \noindent \textbf{C. Offline ILP and \syspol Prefer Higher Accuracy under Low Load.}
    We make yet another observation from the latency profiles of sampled pareto-optimal subnets in \figref{fig:lat_profile:heatmap}. For a batch size $B$, such that $B = B_1 + B_2$ where $B_1 > B_2$, the following holds true in many cases - 
   $B_1 \cdot \text{Acc}(\phi_{high}) + B_2 \cdot \text{Acc}(\phi_{low}) > B \cdot \text{Acc}(\phi_{mid})$. Therefore, $\mathbb{U}(\phi_{high},B_1, d_q) + \mathbb{U}(\phi_{low},B_2, d(B_2)) \geq \mathbb{U}(\phi_{mid},B,d_q)$, may hold true under low load, where the most urgent query q in batch $B$ can be served by either a) mid accuracy model ($\phi_{mid}$) with batch size $B$, or b) high accuracy model ($\phi_{high}$) with larger-size batched partition $B_1$ ($q \in B_1$) with rest of the queries in batch $B_2$ served  with the low accuracy model ($\phi_{low}$) and meeting deadline $d(B_2)$. In such cases, ILP will tend to option (b) \ie an option with higher average accuracy. \syspol also tends to higher accuracy subnets under lower load, as described in \secref{sec:pol:slackfit}.

\section{\system: System Architecture}

\label{sec:sysarch:arch}
\begin{figure}[t!]
	\centering
    \includegraphics[scale=0.6, width=\columnwidth]{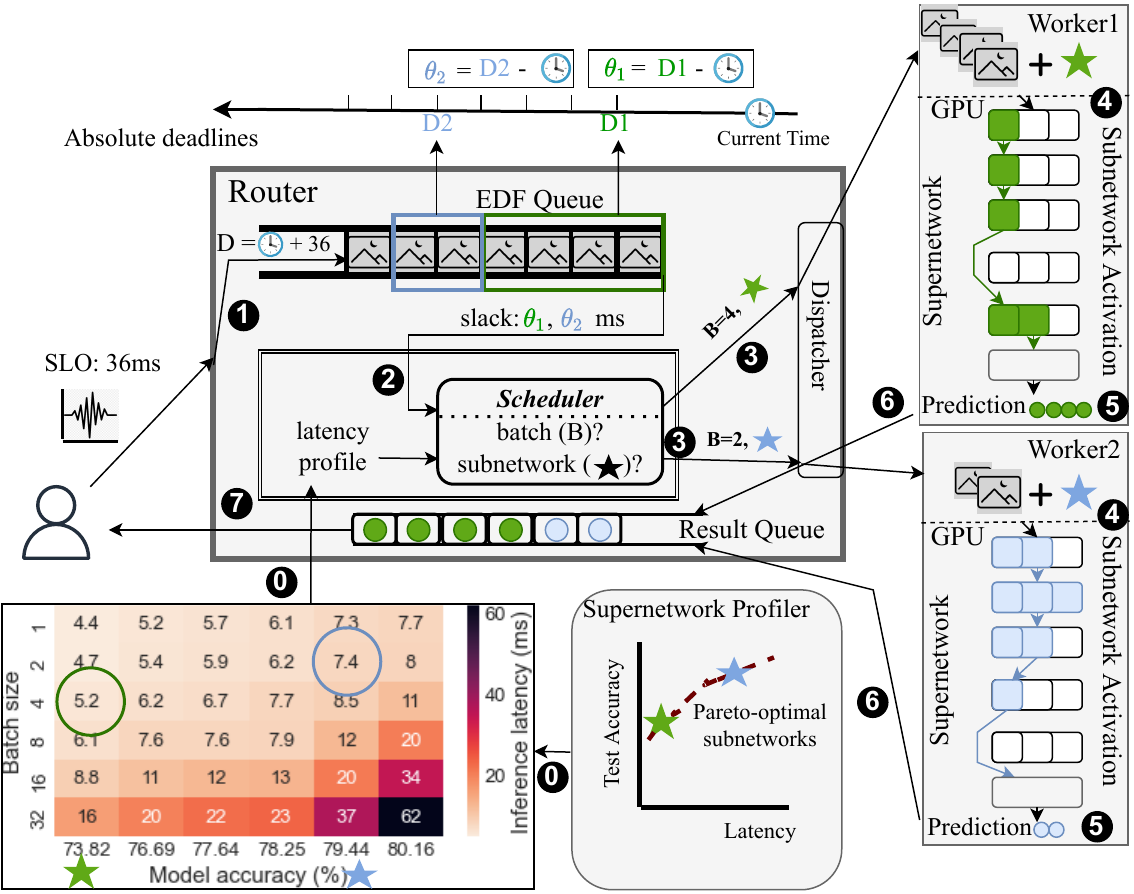}
    \caption{\small 
        \textbf{\system's Architecture.} 
        \system consists of a router, scheduler, workers and a supernetwork profiler. 
        Clients submit queries with a specified latency constraint (SLO) asynchronously. 
        \ding{182} - \ding{188} represents the critical path of the query in \sysname. 
        The latency profile estimation is done apriori, off the critical path by the profiler. 
    }
	 \label{fig:sys_arch_detailed}
	 \vspace{-0.2in}
\end{figure}

\system is a system that instantiates both \sysmech mechanism and \syspol policy. \system's architecture is illustrated in \figref{fig:sys_arch_detailed}. 
The \system client submits asynchronous RPC queries to the router with a deadline. 
These queries are enqueued to a global earliest-deadline-first (EDF) queue (\ding{182}). As soon as any worker becomes available, \system's fine-grained scheduler is invoked (\ding{183}). It decides on the query-batch ($B$) and the subnet ($\phi$)  which are then dispatched to the worker (\ding{184}). Upon receiving this query-batch, the worker that instantiates the supernet instantaneously actuates the chosen subnet in-place on the GPU using \sysmech (\ding{185}), performs inference (\ding{186}), and returns predictions for the query-batch (\ding{187}). The router redirects these predictions back to the client (\ding{188}). 
\par \noindent{\bf Router. }
\label{sec:sysarch:arch:router}
The router runs fine-grained scheduler and interfaces with workers via RPCs.
All queries are received, enqueued, and dequeued asynchronously in the router. It maintains pending queries in a global EDF queue, ordered by timestamps which denote query deadlines. 
The router invokes the scheduler whenever (a) a worker signals availability and (b) the EDF queue is not empty.
It then sends query-batches to workers and also passes back the predictions to the clients.

\noindent{\bf Fine-grained Scheduler. }
\label{sec:sysarch:arch:sched}
The scheduler's control decision is a batch-size and subnet ($\phi = (\mathbb{D}, \mathbb{E}, \mathbb{W})$). 
The scheduler provide pluggable APIs for different policy implementations. \syspol is one such policy implemented in the scheduler. 
All policies in scheduler leverage two key properties to make control decisions: 
(a) predictability of DNN inference latency, %
(b) fast actuation of \sysmech on the query's critical path.

\noindent{\bf Worker. }
\label{sec:sysarch:arch:worker}
The DNN worker employs the \sysmech mechanism to host a supernetwork (\textbf{R3}). \sysmech's operators are implemented in  TorchScript's intermediate representation (IR) \cite{torchscript}.
After receiving a query-batch and subnet ($\mathbb{D}$, $\mathbb{E}$, $\mathbb{W}$) from the router, the worker actuates the desired subnet inplace using \sysmech. 
A forward pass on the actuated subnet produces predictions that are returned to the router. The router's scheduler gets notified about worker availability on receiving the predictions.

\noindent{\bf Supernet Profiler. }
\label{sec:sysarch:arch:profiler}
A supernet profiler is used when a supernet is submitted to \system. This profiling completes apriori, before the workload begins, and off the critical path of the queries. 
The profiler first employs neural architecture search (NAS) \cite{ofa} to find pareto-optimal subnetworks from the supernetwork for each latency target (key design choice of \syspol \secref{sec:pol:slackfit}).  
The latency profiling is done the pareto-optimal subnets obtained by NAS. 
This latency is a function of batch size and target worker GPU (latency profile in \figref{fig:sys_arch_detailed} on RTX2080Ti GPU), and the profiling process for the subnetworks is no different than the model profiling done for other existing models like ResNets \cite{resnet}, Wide-ResNets \cite{wide_resnet}, ConvNeXt \cite{convnext} etc.

\label{sec:fine-grained-scheduler}

\section{Evaluation}
\label{sec:eval}

We assess \system's end-to-end performance \ie its ability to maximize SLO attainment (\textbf{R1}) and accuracy (\textbf{R2}) under a variety of traffic conditions, including synthetic traces (\secref{sec:expt:eval:synthetic}) and real-world derived Microsoft Azure Functions trace (\secref{sec:expt:eval:real}). \system is resource-efficient (\textbf{R3}) due to the use of \sysmech mechanism, already established in \secref{sec:mech:micro}. We conclude with microbenchmarks (\secref{sec:expt:eval:micro}) that show \system linearly scaling to 33,000 qps and providing transparent fault tolerance.
\subsection{Experimental Setup}
\noindent \textbf{Success metrics.}
\textit{SLO attainment} is defined as the fraction of the queries that complete within the latency deadline (\textbf{R1}). The \textit{mean serving accuracy} is calculated for the queries that satisfy the SLO and is the average of models' profiled accuracy that were used to serve the queries (\textbf{R2}).

\noindent \textbf{Traces.}
We evaluate \system on three sets of traces: bursty, time-varying, and real-world. Bursty and time-varying traces are synthetic, similar to those used in InferLine~\cite{inferline}. 
We construct the \textit{bursty traces} by starting with a base arrival with mean ingest rate $\lambda_b$ (with $CV^2=0$) and add a variant arrival trace with mean ingest rate $\lambda_v$ drawing inter-arrival times from a gamma distribution (\figref{fig:expt:syn:sys_dyn:gamma}).
We vary $\lambda_b$, $\lambda_v$ and $CV^2$.
\textit{Time-varying} traces differ from bursty by varying the mean ingest throughput over time. 
We change the mean from $\mu=1/\lambda_1$ to $\mu=1/\lambda_2$  at rate $\tau$ $q/s^2$ with a fixed $CV^2_a$.
Higher ingest acceleration $\tau$ $q/s^2$ 
corresponds to faster change from $\lambda_1$ to $\lambda_2$.
All synthetic trace generation is seeded.
Lastly, we use a MAF trace~\cite{maf} for evaluation on a real-world workload.

\noindent \textbf{Baselines.} 
We compare \system with the single model serving systems that don't perform accuracy trade-offs (and the models are manually selected by users, non-automated serving systems in \secref{sec:prevwork}). These systems are represented as Clipper$^{+}$ baseline and include systems like Clipper \cite{clipper}, Clockwork \cite{clockwork}, and TF-serving \cite{tfserving}. Clipper$^{+}$ is manually configured to serve six different accuracy points (subnets) that uniformly span the supernet's accuracy range and result in its six different versions.
We also compare \system with INFaaS and
note that 
INFaaS is designed to ``pick the most cost-efficient model that meets the [specified] accuracy constraint''~\cite{infaas-fixedacc,infaas,infaas_code}.
However, in the presence of unpredictable, bursty request rates, the choice of the model
accuracy to serve in order to meet the SLO requirements is unknown.
Since, unlike \system{}, INFaaS does not automatically discover the accuracy of the model
to serve under unpredictable request rates and instead requires queries to be hand-annotated
with accuracy thresholds, we choose to run INFaaS with no accuracy thresholds provided (\secref{sec:expt:eval:synthetic:burst},\secref{sec:expt:eval:synthetic:tau}).
In such a scenario, INFaaS reduces to serving the most cost-efficient model (which
is the model with the minimum accuracy).
We confirmed this behavior with the INFaaS authors, who agree that
``[our] representation of INFaaS as a baseline that always chooses the same model is correct in the absence of an accuracy threshold, or a fixed (never changing) accuracy threshold.''~\cite{infaas-fixedacc}. 

\noindent \textbf{Subnet-Profiling.}
We use the supernet trained on ImageNet \cite{imagenet} dataset released by \cite{ofa} and enable \sysmech in it. We extract pareto-subnets ($\Phi_{\text{pareto}}$) by running NAS (publicly released by \cite{ofa}) on trained supernet.
The pareto-subnets in the supernet span $0.9-7.5$ GFLOPs range and an accuracy range of $73-80\%$.
Pareto-subnets are profiled with varied batch sizes on NVIDIA RTX2080Ti GPU. 

\noindent \textbf{Test bed.}
\system is implemented in C++ (17,500 lines of code). gRPC \cite{grpc} is used for communication between a client, the router and workers. 
The experiments use 8 RTX2080Ti GPUs and 24 CPU cores. Each worker uses one GPU.
\label{sec:expts:setup}

\begin{figure*}[t!]
	\centering
    \includegraphics[width=\textwidth]{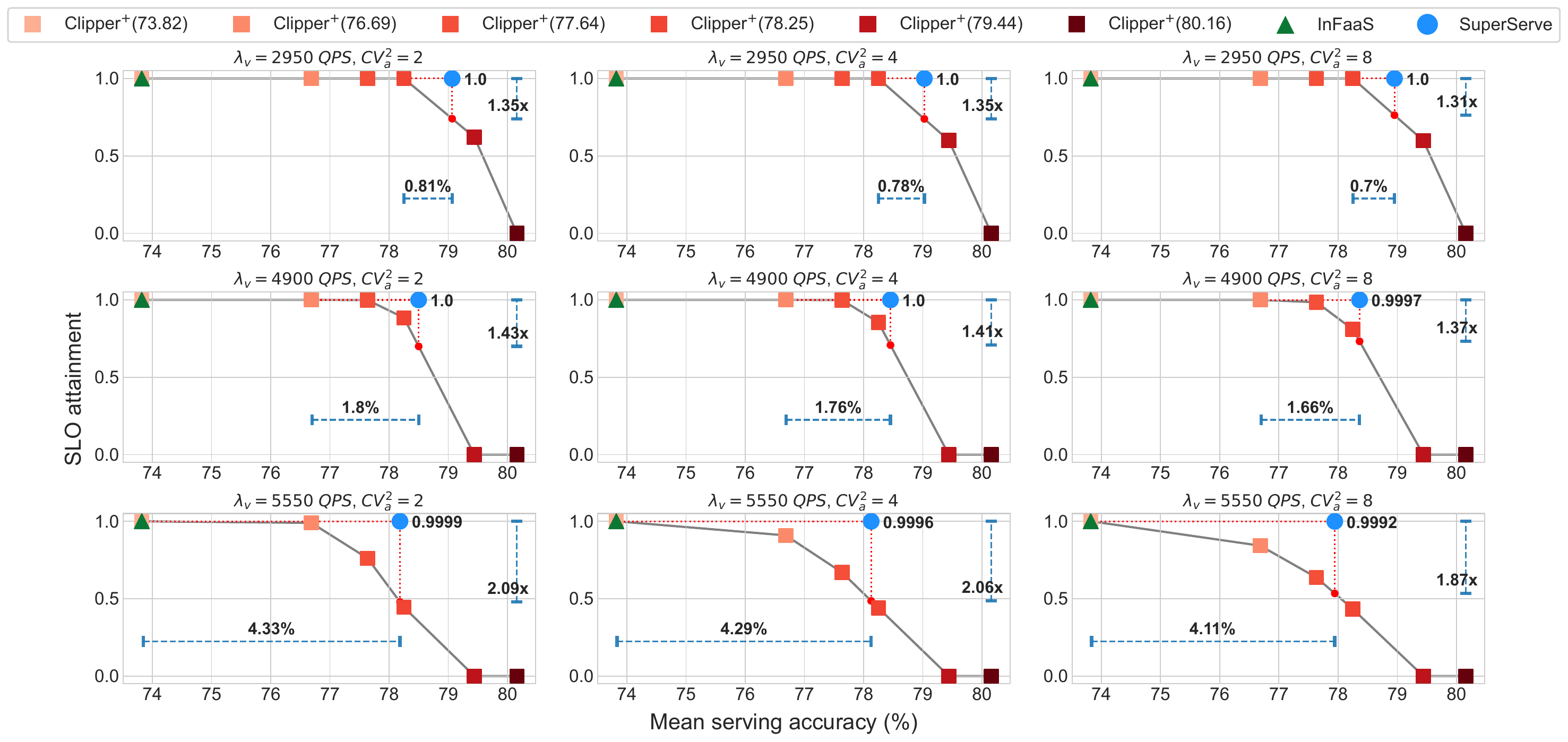}
    \caption{%
        \textbf{\system with variable burstiness.} 
        \system outperforms Clipper$^{+}$ and INFaaS baselines by finding better tradeoffs and consistently achieving $>0.999$ SLO attainment  on bursty traces.
        Variable ingest rate  $\lambda_v=\{$ $2950,$ $4900,$ $5550\}$ q/s increases vertically (down).
        $CV^2_a=\{2,4,8\}$ increases horizontally (across).
        \system achieves a better trade-off in SLO attainment (y-axis) and mean serving accuracy (x-axis) in all cases.
        \system consistently achieves high SLO attainment $>0.999$. 
        }
	 \label{fig:expt:burst}
\end{figure*}

\begin{figure*}[t!]
	\centering
    \includegraphics[width=\textwidth]{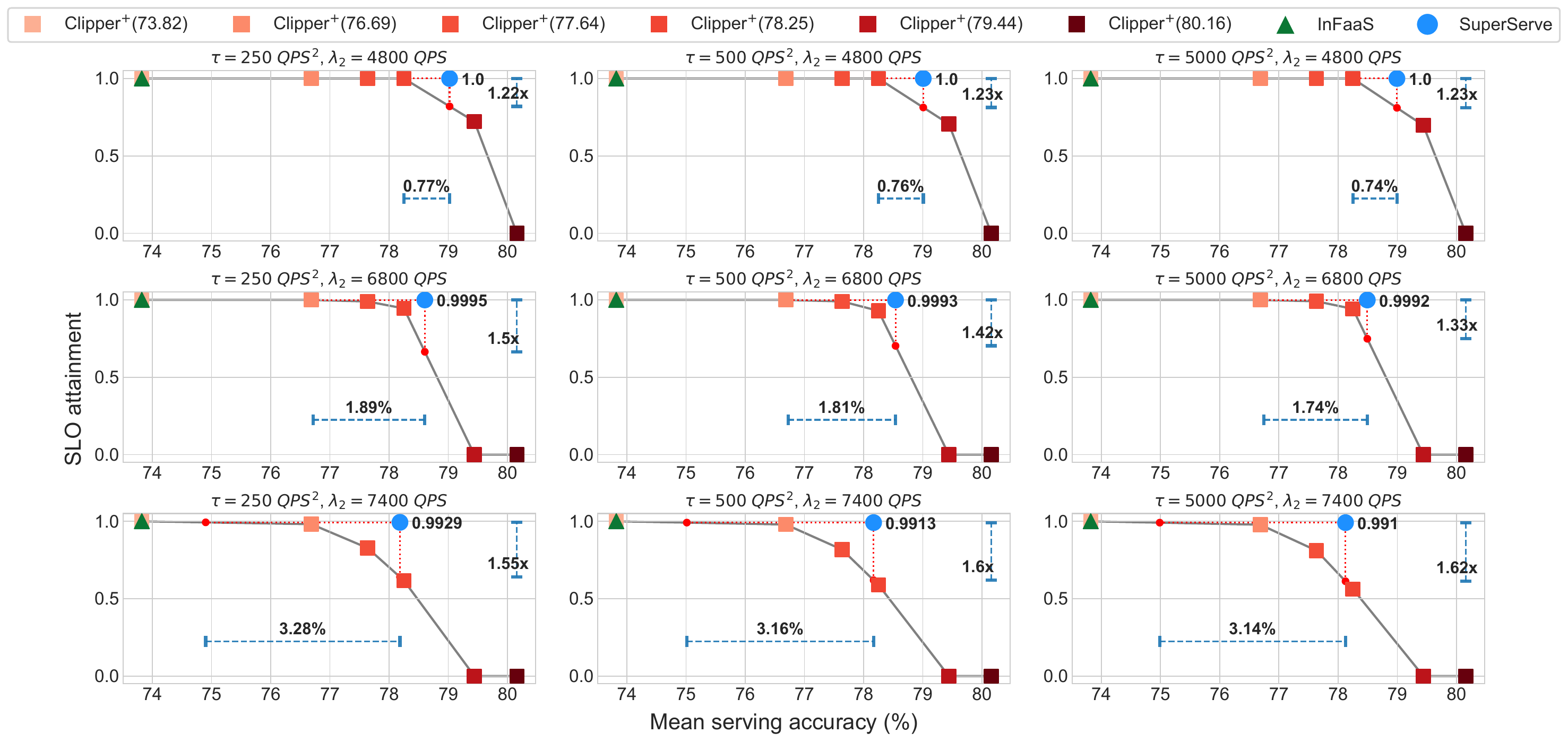}
    \caption{%
        \textbf{\system with arrival acceleration.} 
        \system outperforms Clipper$^+$ and INFaaS baselines by finding better tradeoffs on time varying traces. 
        Mean ingest rate accelerates from $\lambda_1$ to $\lambda_2$ q/s with $\tau$ $q/s^2$. 
        $\tau=\{250, 500, 5000\}$ increases horizontally (across), while $\lambda_2=\{4800, 6800, 7800\}$ increases vertically (down) with $\lambda_1=2500$ q/s  and $CV^2_a=8$ staying constant.
        \system finds a better trade-off in SLO attainment (y-axis) and mean serving accuracy (x-axis). 
    }
	 \label{fig:expt:tau}
\end{figure*}

\subsection{End-to-End: Synthetic}
\label{sec:expt:eval:synthetic}
We aim to answer the following questions, whether \system (a) \textit{automatically} serves queries using appropriate models (accuracy) for different traces (\textbf{R2}),  (b)  achieves a better trade-off \wrt the success metrics (\textbf{R1-R2}),  (c) withstands sharp bursts while maintaining high SLO attainment (\textbf{R1}) and (d) \textit{instantaneously} changes system capacity to serve traces where mean ingest rate changes over time. To answer these questions, we evaluate \system on the bursty and time-varying traces (\secref{sec:expts:setup}).

\subsubsection{Baseline comparison with burstiness}
\label{sec:expt:eval:synthetic:burst}
\figref{fig:expt:burst} compares \system with the baselines over a range of traces increasing 
mean ingest rate $\lambda_v$ across and $CV^2_a$ down.
All traces are configured with $36$ms SLO. 
Achieving high SLO attainment (\textbf{R1}) and high mean serving accuracy (\textbf{R2}) is desirable, which implies the \textit{best trade-off is in the top-right corner of the graph}. 
We demonstrate that no single choice of a model is \textit{sufficient} for different mean arrival rates and $CV^2_a$. 
For instance, the SLO attainment of Clipper$^{+}$(76.69)  decreases as the $CV^2_a$ increases for $\lambda_v=5550$ (row 3). 
Similarly, the SLO attainment of Clipper$^{+}$(78.25) decreases with increase in $\lambda_v$ for $CV^2_a=2$ (column 1). 
We draw the following takeaways:
    \textbf{(1)} \system achieves a significantly better trade-off between SLO attainment and accuracy (\textbf{R1-R2}) than the baselines(Clipper$^{+}$ and \infaas). It is $4.33\%$ more accurate than the baselines at an SLO attainment level of $0.9999$ and  $2.06$x higher SLO attainment at the same accuracy level. 
    \system is consistently at the top-right corner in \figref{fig:expt:burst} across all the traces. 
    \textbf{(2)} \syspol \textit{automatically} selects appropriate models for sustaining different traffic conditions. 
    As $\lambda_v$ increases, \system reduces serving accuracy while maintaining high SLO attainment (columns).

Note that, across all the traces, \infaas achieves an optimal SLO attainment but with a significantly smaller mean serving accuracy (by up to $4.33\%$) than \system. \infaas policy serves the min-cost (and hence min accuracy) model for the trace without accuracy constraints. 
Whereas, \system achieves a better trade-off between the success metrics because 
(a) \sysmech allows in place activation of different subnetworks \textit{without} affecting SLO attainment (\textbf{R1});
(b) \syspol opportunistically selects higher accuracy models based on query's slack (\textbf{R2}). 
Also, the difference between \system and Clipper$^{+}$ narrows \wrt accuracy as $CV^2_a$ increases. This is because \syspol switches to lower accuracy models more frequently with burstier traffic. 
This system dynamics is detailed in \secref{sec:expt:eval:synthetic:sys_dyn}.

\subsubsection{Baseline comparison with arrival acceleration}
\label{sec:expt:eval:synthetic:tau}
\figref{fig:expt:tau} evaluates \system performance at different levels of arrival rate change (i.e., arrival \textit{acceleration}). Traces start at $\lambda_1$ and increase to $\lambda_2$ with acceleration $\tau$.
Traces fix $\lambda_1=2500 qps$ and $CV^2_a = 8$ but change $\lambda_2$ 
and acceleration $\tau$ .

The $\tau$ and $\lambda_2$ are chosen to demonstrate that single, pre-configured model choices are inadequate to sustain different rates of arrival (mean $\lambda$) and acceleration ($\tau$). Clipper$^{+}$($79.44$) starts diverging  as $\tau$ increases ( $\lambda_2$ is $6800$ $qps$ (row 2)). 
Similarly, Clipper$^{+}$($79.44$) starts diverging with increase in $\lambda_2$ ($\tau=250$ $q/s^2$ (column 1)). 
The key takeaways from this experiment are as follows:
\begin{tightitemize}
\item \system \textit{rapidly} scales system capacity  and achieves a high SLO attainment ($0.991$-$1.0$) even with high values of  $\tau$ ($5000$ $q/s^2$).
The experiment demonstrates two key properties of \system---
(a) the \textit{actuation delay} in \system is indeed negligible, 
(b) the lower \textit{actuation delay} helps achieve higher SLO attainments for time-varying traces (\textbf{R1}). 
\system empirically demonstrates ``agile elasticity'' (\secref{sec:back}), and withstands high acceleration in arrival rate ($\tau$).
\item \syspol \textit{dynamically} adjusts the serving accuracy over time (\textbf{R2}) and achieves a better trade-off between success metrics (\textbf{R1-R2}). When the mean ingest throughput is low ($\lambda_1$), \system uses higher accuracy models. It quickly switches to lower accuracy models when mean arrival rate is high ($\lambda_2$), as evident in system dynamics \figref{fig:expt:syn:sys_dyn:tau}.
\end{tightitemize} 

\figref{fig:expt:tau} experiments exhibit interesting trends. As the $\tau$ increases, the gap between \system and Clipper$^{+}$ \wrt mean serving accuracy narrows. This is because \syspol selects smaller accuracy sooner with the increase in $\tau$. 
Lower $\tau$ values give enough time to \system to serve intermediate mean arrival rates with higher accuracy models while gradually moving to lower accuracy models as mean ingest rate increases to $\lambda_2$ qps. 
Whereas, \infaas continues to serve min accuracy model for all traces as its policy doesn't maximize accuracy by design.

\subsection{End-to-End: Real Workloads}
\label{sec:expt:eval:real}

\begin{figure}[t!]
	\centering
	\begin{subfigure}[b]{\columnwidth} %
        \includegraphics[width=\columnwidth, height=0.5\textwidth]{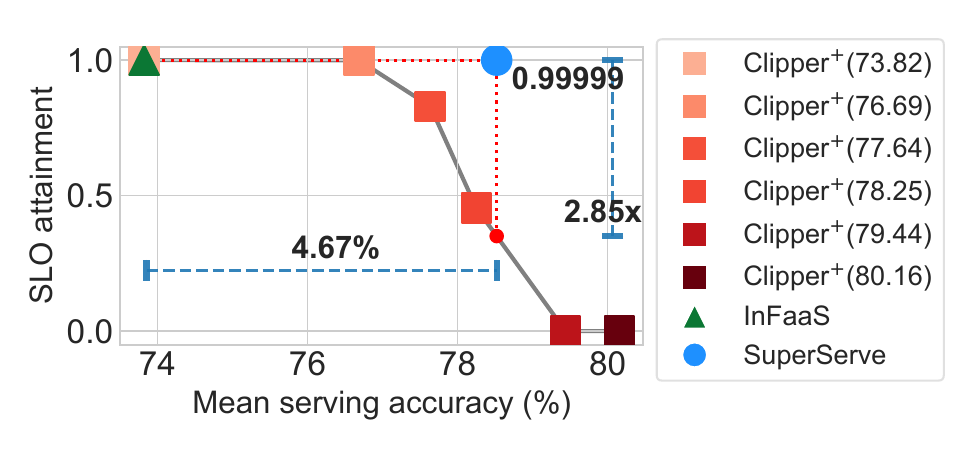}
	    \caption{
		     \system achieves best tradeoffs w.r.t. SLO attainment and accuracy on a real trace.
		}
	    \label{fig:expt:maf:compare}
	\end{subfigure}
	\begin{subfigure}[b]{\columnwidth} %
        \includegraphics[width=\columnwidth, height=0.5\textwidth]{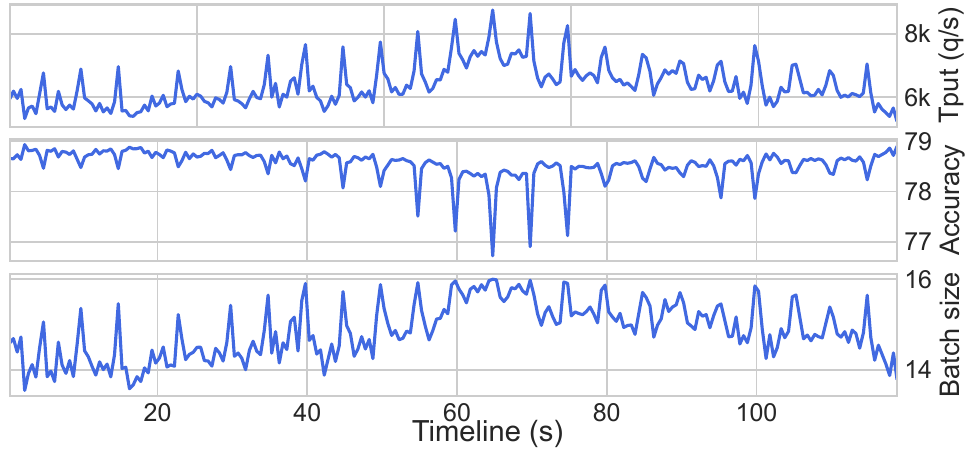}
	    \caption{
		    \system System Dynamics 
		}
	 \label{fig:expt:maf:sys_dyn}
	\end{subfigure}
    \caption{\small 
        \textbf{\system on Real World Trace.} \system on Microsoft Azure Functions (MAF) \protect\cite{maf} trace. 
        (a) \system is compared with Clipper$^+$ and INFaaS baselines, reaching 
         $4.67\%$ higher accuracy at same SLO attainment and $2.85$x higher SLO attainment at same accuracy than any fixed accuracy point that can be served by INFaaS (in the absence of accuracy constraints) and Clipper$^+$. 
        (b) System dynamics w.r.t.  batch size and subnet activation control decisions over time in response to ingest rate in the top graph.
    }
	 \label{fig:expt:maf}
	 \vspace{-0.2in}
\end{figure}
We investigate if: 
(a) \syspol is capable of achieving a better trade-off between SLO attainment and mean serving accuracy on real workloads (\textbf{R1-R2}), and 
(b) \sysmech contributes to serve highly unpredictable workloads at high SLO attainment.

We use the MAF trace \cite{maf} to evaluate \system (similar to Clockwork~\cite{clockwork}). The trace is collected on Microsoft's serverless platform and
serves as a reasonable workload to evaluate \system as serverless ML inference is an active research area \cite{infless,serverless_inference}.
It consists of number of invocations made for each function per minute and contains nearly 46,000 different function workloads that are bursty, periodic, and fluctuate over time. 
We use 32,700 function workloads from the MAF trace, resulting in a mean arrival rate of 6400 qps. The 24 hour long trace is shrunk to 120 seconds using \textit{shape-preserving} transformations to match our testbed.

\myparagraph{Result} \figref{fig:expt:maf:compare} compares \system with Clipper$^{+}$ and \infaas on the real-world MAF trace. 
\system achieves an SLO attainment (\textbf{R1}) of $0.99999$ (five '9's). Compared to Clipper$^{+}$ and \infaas, \system is $4.65\%$ more accurate (\textbf{R2}) at the same level of SLO attainment. It also achieves $2.85$x factor improvement in SLO attainment at the same mean serving accuracy. 
Moreover, Clipper$^{+}$($79.44$, $80.16$) diverges on the MAF trace.

\myparagraph{System Dynamics}
\figref{fig:expt:maf:sys_dyn} shows the ingest throughput (qps), serving accuracy and batch size control decisions (made by \syspol) for the MAF trace. As seen in the figure, the trace contains periodic short-interval spikes that reach upto 8750 qps, demonstrating the agility of the system. 
\syspol selects both smaller accuracy model and higher batch size during the load spikes to meet the deadline (\textbf{R1}). \syspol makes such control decisions because it uses query's slack as a signal to maximize batch size. As the query slack decreases, it selects  maximum batch size control parameters in the lower latency buckets. Furthermore, these control decisions increase the system capacity instantly through \sysmech. Lastly, \syspol serves higher accuracy models when the ingest rate is low and hence, achieves better 
mean serving accuracy (\textbf{R2}).

\begin{figure*}[t!]
    \centering
    \begin{subfigure}[b]{0.33\textwidth}
        \includegraphics[width=1\textwidth,keepaspectratio]{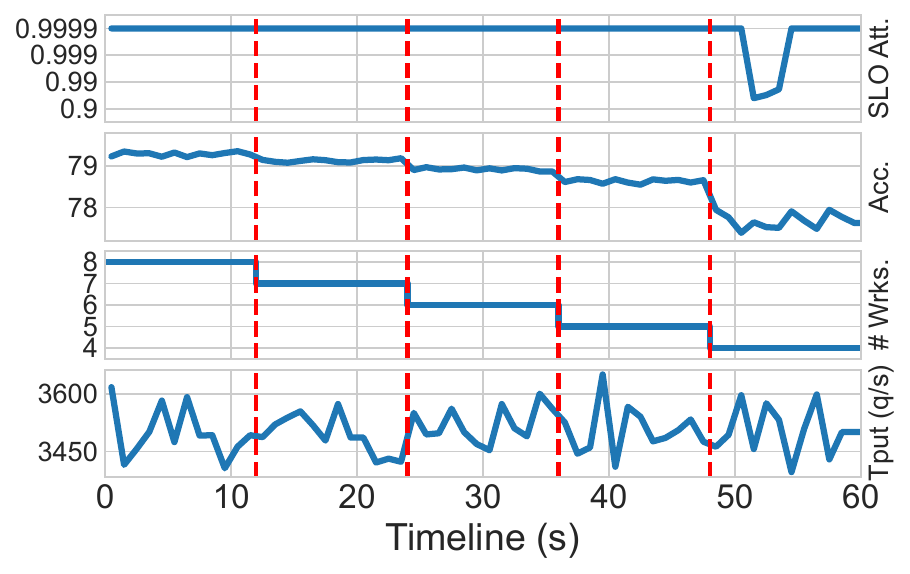}
        \caption{\small Fault-Tolerance}
         \label{fig:expt:ft}
         \label{fig:expt:micro:ft}
         \vspace{-0.1in}
    \end{subfigure}%
    \hfill
    \begin{subfigure}[b]{0.33\textwidth}
        \includegraphics[width=1\textwidth,keepaspectratio]{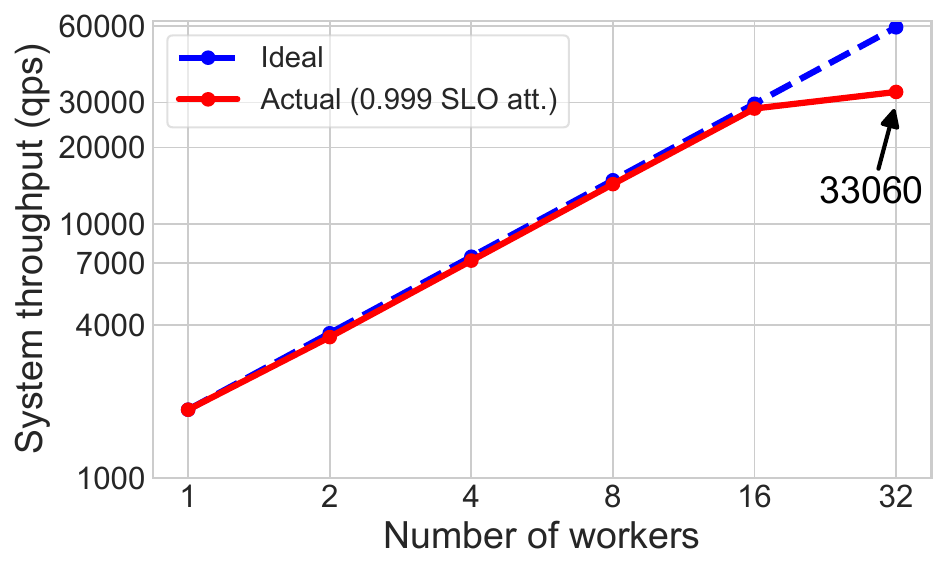}
        \caption{\small Scalability}
         \label{fig:expt:scale}
         \label{fig:expt:micro:scale}
         \vspace{-0.1in}
    \end{subfigure}%
    \hfill
    \begin{subfigure}[b]{0.33\textwidth}
        \includegraphics[width=1\textwidth,keepaspectratio]{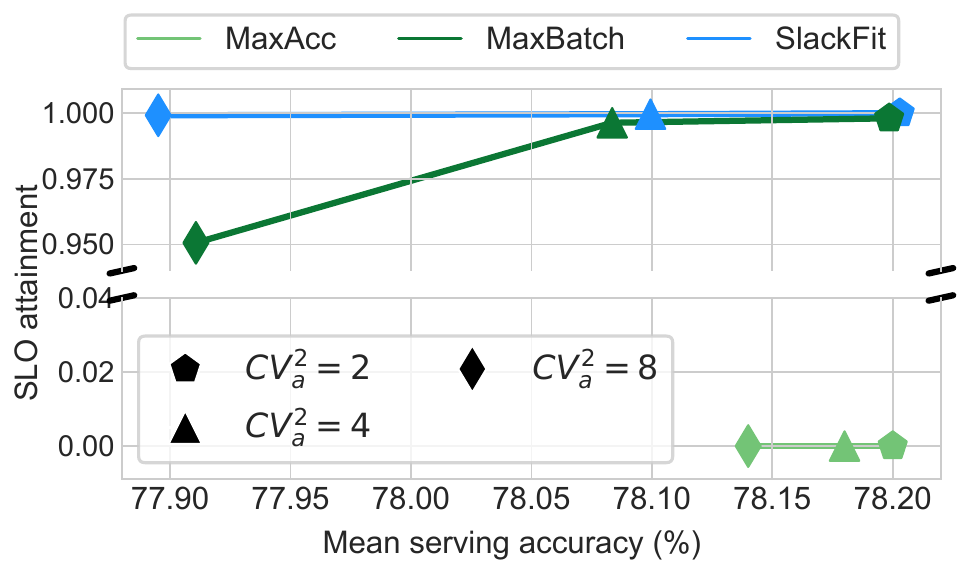}
        \caption{\small Policy-SE}
         \label{fig:expt:policy_micro}
         \label{fig:expt:micro:policy}
         \vspace{-0.1in}
    \end{subfigure}
    
    \caption{\small 
        \textbf{\system's Micro-benchmarks.}
        \textbf{(a)} \system resiliency to faults. \system maintains high SLO attainment in the system by dynamically adjusting served accuracy as workers drop out over time. The trace stays statistically the same ($\lambda=3500$ qps, $CV^2_a=2$ (last row)).
        \textbf{(b)} \system scales linearly with the number of workers, achieving up to 33000 qps (orders of magnitude higher than published SotA systems) while maintaining high .999 SLO attainment. 
        \textbf{(c)} \syspol finds the best tradeoff on the SLO attainment/accuracy maximization continuum automatically (\secref{sec:expt:eval:micro:policy}).
    }
    \label{fig:expt:micro}
\end{figure*}

\subsection{Microbenchmarks}
\label{sec:expt:eval:micro}
\label{sec:expt:eval:micro:ft}
{\bf Fault Tolerance.}
\sysmech mechanism provides an additional advantage of transparent fault tolerance.
We run \system 
with $100\%$ capacity ($8$ workers) with a bursty traffic trace ($\lambda=3500$ qps, $CV^2_a=2$) for $60$ seconds and gradually kill a worker every $12$ seconds to simulate faults.
\system shows resilience to decreases in system throughput capacity to as low as 50\% by maintaining SLO attainment as high as 0.999 for the unchanging trace 
as it leverages subnetwork activation to serve lower accuracy models automatically. 
Similar methodology was used in \cite{lineage_stash}.
\figref{fig:expt:ft} shows SLO attainment as a function of time (along with other system dynamics). As the faults occur (workers killed, dotted red lines), 
\system \textit{automatically} transitions to lower accuracy models to maintain high SLO attainment. 
We attribute \system's fault tolerance to 
(a) a wide-dynamic throughput range afforded by \sysmech (\figref{fig:subnetact:benefit:th_range}) that allows \system to serve the workload even with $50\%$ capacity, and 
(b) \sysmech's low actuation delay that provides \textit{agility} to rapidly increase system-capacity (during faults) without sacrificing SLO attainment (\textbf{R1}).

\noindent{\bf Scalability.}
\label{sec:expt:eval:micro:scale}
We assess if \system reaches high SLO attainment at scale. To show this, we scale the number of workers and observe the maximum throughput \system sustains to reach SLO attainment of $0.999$. We serve Resnet-18 \cite{resnet} across all the workers with clients providing a batch of 8 images\footnote{we don't perform adaptive batching for this experiment}. Scalability experiments are conducted with $CV^2_a=0$. \figref{fig:expt:micro:scale} shows sustained ingest throughput with the increase in workers. In this experiment \system achieves an SLO attainment of $0.999$ while reaching throughputs as high as $\approx33000$ qps.

\noindent{\bf Policy Space Exploration.}
\label{sec:expt:eval:micro:policy}
We compare different policies implemented in \system (\figref{fig:expt:policy_micro}). We show that \syspol achieves the best tradeoff \wrt our success metrics compared to both MaxAcc (greedily maximizes accuracy) and MaxBatch (greedily maximizes throughput) as $CV^2_a$ is varied. Details of the policies and the experiment are in \secref{sec:sched_policies:impl}.

\section{Related Work}
We build on recent DNN supernet training progress, which is complementary to this work. 
\cite{ofa,compofa} train supernets for image classification vision tasks~\cite{image-classification}. 
Dynabert~\cite{dynabert} supernet supports NLP tasks trained on  GLUE \cite{glue}, including textual entailment, question answering and sentiment analysis. 

\noindent \textbf{Model serving systems} can be broadly divided into two categorizes --- a) Non-Automated, and b) Automated. \textit{Non-automated serving system} expect application developers to provide the prediction models and make explicit choices in the accuracy-latency trade-off space. This category includes TensorFlow serving~\cite{tfserving}, Clipper~\cite{clipper}, InferLine~\cite{inferline}, SageMaker \cite{sagemaker}, Triton~\cite{triton}, Shepherd \cite{zhang2023shepherd} and Clockwork \cite{clockwork}. TensorFlow Serving serves the models trained in TensorFlow framework while Clipper and Triton support  models trained from multiple frameworks. Triton optimizes models for GPU serving. Clockwork guarantees predictable tail latency for DNN inference by designing a predictable system bottom up and making cross-stack design decisions explicitly for worst case predictability. Inferline provides support for provisioning inference pipelines that consist of multiple models, but the models are still hard coded in the pipeline vertices. 
However, all the prior works in this category are complementary to \system. For instance, \system's workers can be made more predictable by consolidating choices like Clockwork. Inferline's autoscaling policy can be used on top of \system. The max-provisioning ratio of the smallest subnetwork in \system can be calculated to construct traffic envelopes for Inferline's high frequency tuner. The workers can then be scaled up if the ingest throughput cannot be served by the smallest subnetwork of the supernet. 
\par \noindent In contrast, \textit{automated serving systems}~\cite{infaas,modelswitch} provide a mechanism for switching between available latency/accuracy points and automate the navigation of the accuracy-latency trade-off space with a policy, resulting in automatic DNN selection at runtime. 
However, both \cite{infaas} and \cite{modelswitch} use state-of-the-art DNNs (e.g., ResNets, MobileNets) and rely on model loading mechanism instead of the supernet, which offers better pareto-optimality and orders of magnitude faster model switching enabled via proposed \sysmech. Fundamentally, these state of the art mechanisms implicitly bias their policies to avoid model switching, which clearly limits the ability of the system to respond to unpredictable trace and query complexity dynamics in a agile fashion.
InFaaS's DNN switching policy is biased towards selecting the least accurate DNNs that satisfy accuracy constraints, as the goal of the stated goal of the system is to satisfy constraints instead of treating accuracy as an optimization objective.
Model-Switching~\cite{modelswitch} switches between CPU models only. Thus, it cleverly never incurs the overhead of GPU model loading faced by other systems such as InFaaS and Clockwork. \system supports GPU model serving via subnetwork activation, addresing the model switching overhead through its proposed \sysmech. 

\myparagraph{Training Supernets} Supernet (as a concept) and its training was first proposed by OFA \cite{ofa}. The subnetworks trained as a part of training the supernet are shown to offer a better accuracy-latency trade-off than existing models like EfficientNets \cite{efficientnet}, ResNets \cite{resnet}, MobileNets \cite{mobilenet} and SlimmableNets \cite{slimmable}. Furthermore, there is a surge of recent works that propose improvements to the supernet training such as CompOFA \cite{compofa}, and BigNAS \cite{bignas}. For instance, CompOFA makes the training of Supernets faster and more accurate by training a fewer number of subnetworks simultaneously. While BigNAS trains the supernet in a one-shot fashion with a wider range of subnetworks. Dynabert \cite{dynabert} trains a supernet based on transformer neural network architecture for text datasets. It varies the depth and hidden dimensions in BERT-like models. 
AutoFormer \cite{autoformer} is another technique that trains supernets derived from vision transformers. The subnetworks extracted from AutoFormer's supernet surpass state-of-the-art vision transformers like ViT \cite{vit} and DeiT \cite{deit}. NasViT \cite{nasvit} trains the supernet for semantic segmentation tasks and achieves a better trade-off between  accuracy and latency at fewer FLOPs.
\system provides system support for serving Supernets trained using any existing technique.   
We believe that Supernets are an emerging phenomenon, and system support for serving them is the need of the hour. 
\label{sec:prevwork}

\section{Conclusion}
We describe a novel mechanism \sysmech{} that carefully inserts specialized control-flow and slicing operators
into SuperNets to enable a resource-efficient, fine-grained navigation of the latency-accuracy tradeoff space.
\sysmech{} unlocks the design space of fine-grained, reactive scheduling policies.
We explore the design of one such simple, yet effective greedy heuristic-based cheduling policy \syspol{}.
We instantiate \sysmech{} and \syspol{} in \sysname{}, and extensively evaluate it
on real-world workloads.
\sysname{} achieves $4.67\%$ better accuracy at the same level of SLO attainment or 2.85x better SLO attainment at the same level of serving accuracy compared to state-of-the-art inference-serving systems.
\label{sec:conclusion}

\newpage
\bibliographystyle{plain}
\bibliography{bibs/datacenter,bibs/superserve, bibs/related_work}

\newpage
\appendix
\section{Appendix}
\subsection{Utility of Pareto Points is Higher}
\label{sec:pareto:utility}
\begin{lemma}
    The utility of pareto-optimal subnets is higher than non-pareto-optimal subnets if they have similar inference latency for a batch of queries.
    \begin{equation*}
    \begin{split}
        \mathbb{U}(\phi_p,B,d_B) > \mathbb{U}(\phi_q,B,d_B), \;\;\;\;\;\;\;\;\forall B,d_B \\ 
        \text{s.t.}\;\; \phi_p \in \Phi_{\text{pareto}},\; \phi_q \in \{\Phi \setminus \Phi_{\text{pareto}}\}, \; l_{\phi_p}(B) \approx l_{\phi_q}(B) 
    \end{split}
    \end{equation*}
\end{lemma}

\textit{Proof By Contradiction.} Assume a non-pareto optimal subnet ($\phi_q$)  such that it has higher utility than pareto optimal subnet  ($\phi_p$) for a batch $B$ and $l_{\phi_p}(B) \approx l_{\phi_q}(B)$ \ie $\mathbb{U}(\phi_p,B,d_B) < \mathbb{U}(\phi_q,B,d_B)$.

Now, due to the pareto optimal property $Acc(\phi_p) > Acc(\phi_q)$, this implies $Acc(\phi_p) \cdot |B| > Acc(\phi_q) \cdot |B|$ which implies $\mathbb{U}(\phi_p,B,d_B) \geq \mathbb{U}(\phi_q,B,d_B)$ for any delay $d_B$ as $l_{\phi_p}(B) \approx l_{\phi_q}(B)$. This is contradiction. Hence Proved.

\subsection{System Dynamics: Synthetic Traces}

\label{sec:expt:eval:synthetic:dyn}
\label{sec:expt:eval:synthetic:sys_dyn}
We also derive key observations from the dynamics to understand how \system achieves high SLO attainment and better trade-offs (\textbf{R1-R2}) for synthetic traces. 

\begin{figure}[t!]
	\centering
	\begin{subfigure}[b]{\columnwidth} %
        \includegraphics[width=\columnwidth, height=0.5\textwidth]{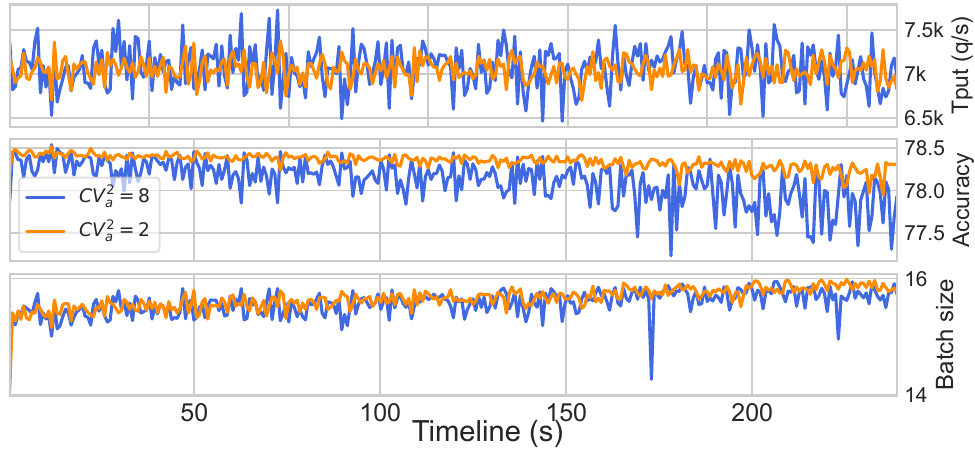}
	    \caption{
		     Dynamic accuracy and batch size control: bursty traces
		}
		\label{fig:expt:syn:sys_dyn:gamma}
	\end{subfigure}
	\begin{subfigure}[b]{\columnwidth} %
        \includegraphics[width=\columnwidth, height=0.5\textwidth]{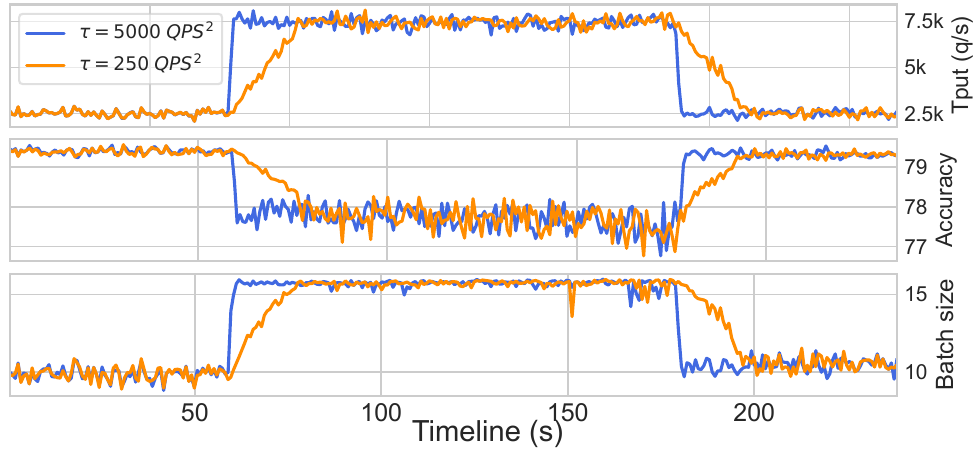}
	    \caption{
		    Dynamic accuracy and batch size control: time-varying
		}
		\label{fig:expt:syn:sys_dyn:tau}
	\end{subfigure}
    \caption{\small 
        \textbf{System Dynamics on Synthetic Traces.} %
            Accuracy and batch size control decisions shown over time in response to ingest throughput (q/s). 
            (a) bursty traces $\lambda = 7000 = (\lambda_b=1500) + (\lambda_v=5500)$ with burstiness of $CV^2_a=2$ (orange) and $CV^2_a=8$ (blue). 
            (b) time varying traces accelerate from $\lambda_1=2500$ q/s to $\lambda_2=7400$ q/s with acceleration $\tau=250 q/s^2$ (orange) and $\tau=5000 q/s^2$ (blue).
            Batch size and subnetwork activation control choices over time show how \system reacts to each of the four plotted traces in real time. This illustrates dynamic latency/accuracy space navigation.%
}
	 \label{fig:expt:syn:sys_dyn}
	 \vspace{-0.2in}
\end{figure}

\figref{fig:expt:syn:sys_dyn} shows the system dynamics of \system for both bursty and time-varying traces. 
The mean ingest rate of the bursty traces is $7000$ qps and they vary in $CV^2_a$ = \{$2$, $8$\}. Similarly, in case of the time-varying traces, the ingest rate is increased from $\lambda_1$ qps to $\lambda_2$ qps at varying accelerations $\tau$ = \{$250$ $q/s^2$, $5000$ $q/s^2$\}. The control decisions made by \syspol (subnetwork (accuracy) and batch size) are shown over time. 

\figref{fig:expt:syn:sys_dyn:gamma} shows system dynamics for the bursty traces. The trace with $CV^2=8$ (blue line) has higher spikes than the trace with $CV^2=2$ (orange line). First, note that \system operates at an accuracy range of $76-78\%$ and never selects a higher accuracy subnetwork such as the subnetwork of $80.16\%$ accuracy. This is because the subnetwork of $80.16\%$ accuracy diverges at the mean ingest rate of $7000$ qps (also seen in \figref{fig:expt:burst} last row). Hence, \system \textit{automatically} selects appropriate subnetworks for different mean ingest rates. Moreover, \system uses lower accuracy models more frequently with increasing $CV^2_a$. This is because increased jitter reduces query slack, causing \syspol to pick lower latency buckets more often. This corroborates the trend seen in \figref{fig:expt:burst} where the mean serving accuracy of \system monotonically decreases as $CV^2_a$ increases. Lastly, during the load spikes, \syspol usually selects control parameters with high batch size and smaller subnetwork (\secref{sec:pol:approx}). This control decision allows \system to drain the queue faster, resulting in a high SLO attainment on the traces (\textbf{R1}). 

\figref{fig:expt:syn:sys_dyn:tau} shows the system dynamics for the time-varying traces. $\tau=5000$  $q/s^2$ (blue line) increases the ingest rate from $2500$ qps to $7400$ qps faster than $\tau=250$ $q/s^2$. For both the traces, \system dynamically changes the accuracy from $\approx79.2$ to $\approx77.5$ as mean ingest rate increases. \system's ability to dynamically adjust accuracy helps it achieve a higher mean serving accuracy (\textbf{R2}) compared to serving a single model statistically. Moreover, for $\tau=5000$ $q/s^2$, \system jumps to lower accuracy and higher batch size control parameters quickly. While, for for $\tau=250$ $q/s^2$, \system uses intermediate models to serve the intermediate ingest rate during $\approx60-80$ seconds. A higher $\tau$ value forces query's slack to reduce drastically. Hence, \syspol rapidly switches to selecting control parameters of smaller subnetwork and higher batch size from the low latency buckets (\secref{sec:pol:slackfit}) to satisfy deadlines (\textbf{R1}). Therefore, increase in $\tau$ decreases mean serving accuracy (a trend observed in \figref{fig:expt:tau} across the rows).

\subsection{Scheduling policies}
\begin{figure}[t!]
	\centering
	\begin{subfigure}[b]{0.3\textwidth}
        \includegraphics[width=\textwidth]{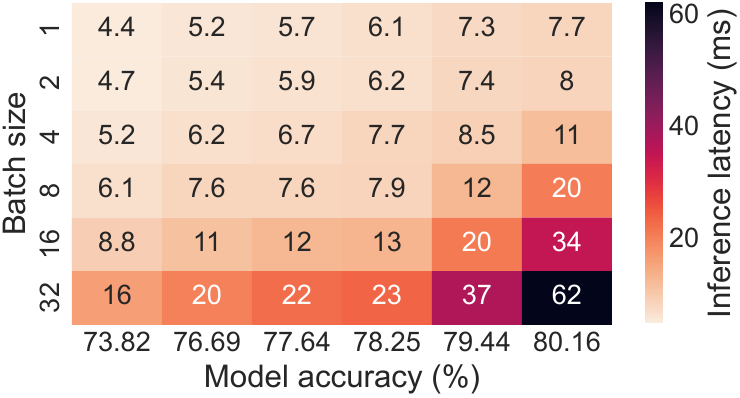}
	    \caption{
		     Latency Heatmap
		}
	\label{fig:lat_profile:heatmap}
	\end{subfigure}
	\begin{subfigure}[b]{0.15\textwidth}
        \includegraphics[width=\textwidth]{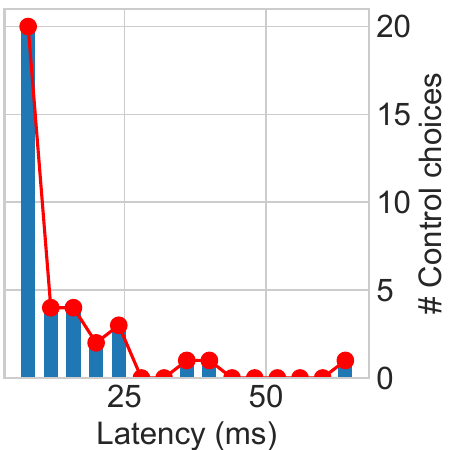}
	    \caption{
		    Control Choices 
		}
	\label{fig:lat_profile:num_choices}
	\end{subfigure}
    \caption{\small
        \textbf{\syspol control parameter space.} 
        (a) latency heatmap for six different (uniformly sampled \wrt FLOPs) pareto-optimal subnets in \sysmech as a function of accuracy (x-axis) and batch size (y-axis). 
        The latency is \textit{monotonic} with batch size and accuracy. 
        (b) the number of available choices
        decrease as latency increases.
    }
	 \label{fig:lat_profile:all}
	 \vspace{-0.2in}
\end{figure}

The core functionality of the \system's scheduler is to maximize 
(a) SLO attainment and 
(b) prediction accuracy for any arrival trace dynamics. %
The scheduler offers a pluggable policy framework to support any application sensitivity to these metrics by allowing arbitrarily different trade-offs between them.
\sysname policy interface dictates that control decisions are made \wrt the batch size and subnetwork to activate.
Both of these control parameters affect SLO attainment and serving accuracy. 
This is because the scheduler a) perpetually operates under a latency constraint and 
b) these control decisions have a cumulative effect over time (e.g., higher accuracy affects queue build-up later). 
Finding a globally optimal set of batch size and subnetwork control tuples over time is NP-hard.
As our control decisions must be made on the critical path of queries' end-to-end latency, 
quick sub-millisecond control decision making is a key performance requirement.
Thus, to meet the real time requirements, we primarily consider scheduling policies that are greedy \wrt time.
The policies decide the batch size and subnetwork based on the remaining slack of the most urgent query. The slack is calculated using a fast (sub-ms) $O(1)$ EDF queue lookup operation.

\subsection{Control Parameter Space}
The control parameter space of the scheduling policies is created using \system's supernetwork profiler (\figref{fig:lat_profile:heatmap}) by profiling latency as a function of subnetwork accuracy and batch size.
With $B$ possible batch size choices and  $S$ different subnetworks to serve the size of the control parameter space  is $B \times S$.
All the scheduling policies use this space to inform their trajectory through the latency/accuracy trade-off space. 

{\bf Insights. }
We draw some key insights from the control parameters' space in \figref{fig:lat_profile:heatmap}.
(\textbf{I1}) the  latency increases monotonically with batch size.
(\textbf{I2}) the  latency increases monotonically with accuracy. 
 (\textbf{I3}) the number of control choices decreases with latency increase (\figref{fig:lat_profile:num_choices})\footnote{this occurs  due to choice of power of 2 batch-size (which are granular enough to observe reasonable latency differences) 
 and availability of fewer model choices at higher latency range.}. 
 These insights inform scheduling policy implementation in in \system. 
 The larger batch size incurs sub-linearly higher latency, which helps increase system throughput.
 Thus, it is always beneficial to maximize batch size subject to the latency constraints. 
 By increasing system throughput, the systems utilization increases, serving more queries in the same amount of time, which helps the SLO attainment. 
 At the same time, a policy may also favor more accurate subnetworks subject to query latency constraints. 
 This increases the overall accuracy of predictions rendered by the system, improving application-visible quality of service.
 Thus, batch size and accuracy forms two levers of control to maximize both SLO attainment and prediction accuracy.

\subsection{Policy Design Space}
\label{sec:sched_policies:impl}

{\bf MaxBatch Policy.}
\label{sec:sched_policies:impl:max_batch}
This policy first maximizes the batch size  and then the accuracy. 
It greedily finds a maximal batch size ($b$) for the smallest accuracy subnetwork that fits within latency slack $\theta$.%
Within the chosen batch size MaxBatch finds the maximum accuracy subnetwork ($s$) such that the profiled latency $L(b,s) < \theta$ .
It returns the control choice  $(b,s)$.
This policy leverages  insights (\textbf{I1}) and (\textbf{I2}). 
It takes O($log(B)$) operations to find $b$ 
and 
O($log(S)$) operations to find $s$ (binary search on  monotonically increasing latency \wrt batch size and accuracy).
As a result, this lightweight policy scales well with the profile table, taking only O($log(B) + log(S)$) operations to make control decisions. 

{\bf MaxAcc Policy.}
\label{sec:sched_policies:impl:max_acc}
MaxAcc first maximizes the accuracy and then the batch size.
Mirroring MaxBatch, MaxAcc
performs a binary search for the largest accuracy ($s'$) with $L(1,s') < \theta$ first. %
Then, it finds the maximal batch size ($b'$) keeping the subnetwork choice fixed to the chosen $s'$, such that
$L(b',s') < \theta$ ms.
Similarly to MaxBatch policy, it leverages  insights (\textbf{I1}) and (\textbf{I2}) and takes O($log(B) + log(S)$) operations to return the control choice $(b',s')$.

{\bf The proposed \syspol Policy.}
\label{sec:sched_policies:impl:bucket}
This is our best performing policy. At a high level, \syspol partitions the set of feasible profiled latencies into evenly sized latency buckets. Each bucket consists of control tuples $(b,s)$ with $L(b,s)$ within the range of bucket width. Then the policy chooses a bucket with latency $\leq \theta$. Finally, from the choices within the selected bucket, it picks the control choice that maximizes batch size.
Intuitively, selecting control parameters closest to slack $\theta$ configures the system to operate as close to capacity as possible.%
In other words, choices with latency less than that either reduce the throughput capacity or the serving accuracy, eventually lowering system's SLO attainment and accuracy. This draws on the monotonicity insights (\textbf{I1}) and (\textbf{I2}).
\syspol's novelty is in insight (\textbf{I3}). We  observe that \syspol \textit{dynamically} detects and adapts to the runtime difficulty of the trace. A well-behaved trace (e.g., low ingest rate, variation, acceleration) results in higher $\theta$. Higher $\theta$ leads to the choice of higher latency buckets. And higher latency buckets  are correlated strongly with fewer control tuple choices (\figref{fig:lat_profile:num_choices}), maximizing the probability of choosing higher accuracy models! 
Conversely, mal-behaved traces (higher ingest rate, variation,  acceleration) lead to lower latency bucket choices, as the scheduler is operating under much lower $\theta$ conditions. There are more control choices in lower latency buckets, which leads to control tuples within those buckets to favor higher batch sizes.
This leads to processing the queue faster!

{\bf Experiment Result.}
In \figref{fig:expt:policy_micro} we show that \syspol achieves the best tradeoff \wrt our success metrics compared to both 
MaxAcc -- a policy that greedily maximizes accuracy and MaxBatch --- a policy that greedily maximizes batches.
The traces used  mean  $\lambda=7000$ qps ($(\lambda_b=1500) + (\lambda_v=5550)$) and $CV^2_a \in \left\{2,4,8 \right\}$.
\syspol reaches the highest SLO attainment($0.999$) for all $CV^2_a$.  
MaxBatch starts under performing \wrt SLO attainment with $CV^2_a$ increase. 
The \syspol and MaxBatch difference is most pronounced at the highest $CV^2_a$, eventually causing 
a significant 5\% drop in the SLO attainment.
Both policies maximize the batch size within latency slack $\theta$ when operating under small $\theta$. 
When $\theta$ increases, however, 
MaxBatch continues to maximize the batch size unconditionally---a greedy choice that leads to packing larger batches. 
This greedy decision causes more time to be spent in a worker compared to \syspol, which adaptively shifts to higher accuracy models under larger $\theta$ conditions with compound effect on queued queries, eventually missing their SLOs.
maxAcc is unable to keep up with this trace. It never switches to policy decisions that process the queue faster. This policy comparison shows a continuum between faster queue processing and serving higher accuracy, with \syspol automatically finding the best point in this continuum.

\section{Serving fixed accuracy points}
\par \textbf{Experiment setup.} The experiments were run for the MAF trace with the mean ingest rate of $4,000$ queries per second. The MAF follows a poisson-like distribution with $CV_a^2=1$ and the latency $SLO$ for each of these requests was set to $30ms$. The scheduler at the router adaptively batches the requests, along with an appropriate model based on the scheduling policy. The query-batch and model configuration is then sent to one of the 6 workers, each of which had an NVIDIA RTX 2080Ti GPU.
\par \textbf{Results.} In the plots, the red solid line represents the latency and accuracy range of all available models. Any query with remaining slack (ms) in the latency range spanned by the models can be served. It must be noted that while the system can serve queries with remaining slack higher than the right-interval red-line latency, queries with remaining slack lower than left-interval red-line latency cannot be serviced and must be dropped. The goal across policies is to manage the queue such that the remaining slack remains within the model latency range, while maximizing accuracy.
\par
\begin{figure*}[!h]
    \centering
    \begin{subfigure}[b]{0.33\textwidth}
        \includegraphics[width=1\textwidth,keepaspectratio]{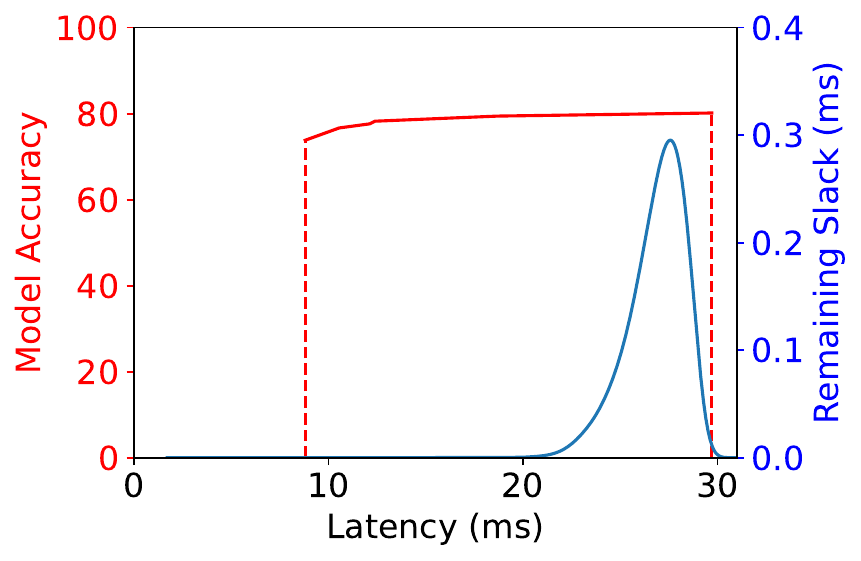}
        \caption{\small Clipper$^{+}$($73.82$)}
         \label{fig:appendix:clipper_73}
         \vspace{-0.01in}
    \end{subfigure}%
    \hfill
    \begin{subfigure}[b]{0.33\textwidth}
        \includegraphics[width=1\textwidth,keepaspectratio]{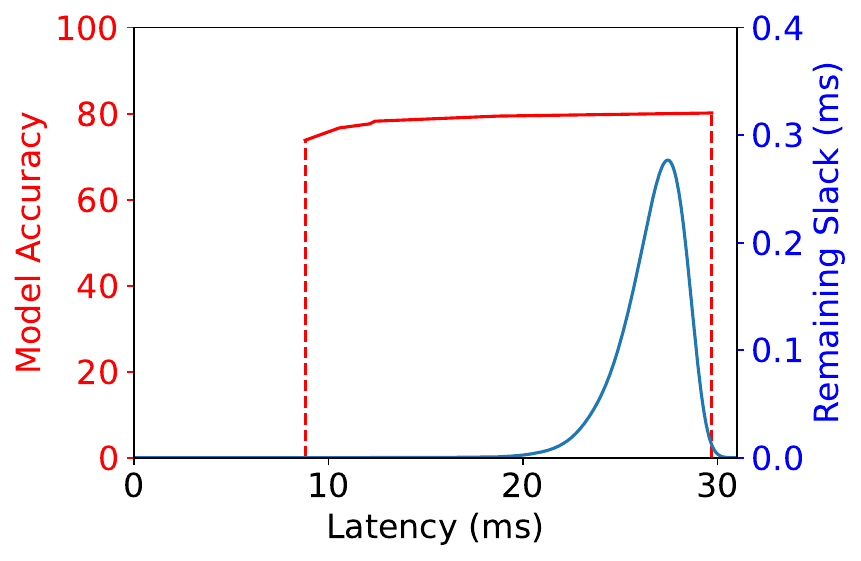}
        \caption{\small Clipper$^{+}$($76.69$)}
         \label{fig:appendix:clipper_76}
         \vspace{-0.01in}
    \end{subfigure}%
    \hfill
    \begin{subfigure}[b]{0.33\textwidth}
        \includegraphics[width=1\textwidth,keepaspectratio]{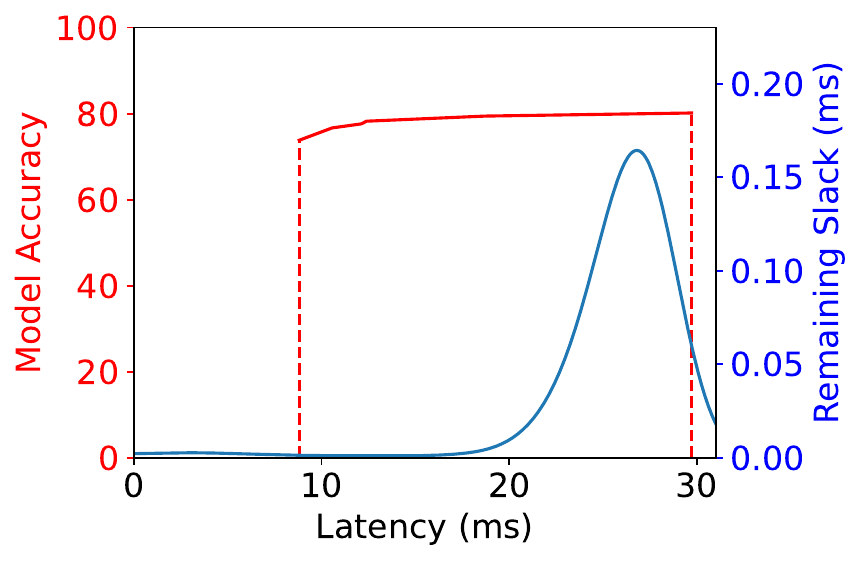}
        \caption{\small Clipper$^{+}$($77.64$)}
         \label{fig:appendix:clipper_77}
         \vspace{-0.01in}
    \end{subfigure}
    \hfill
    \begin{subfigure}[b]{0.33\textwidth}
        \includegraphics[width=1\textwidth,keepaspectratio]{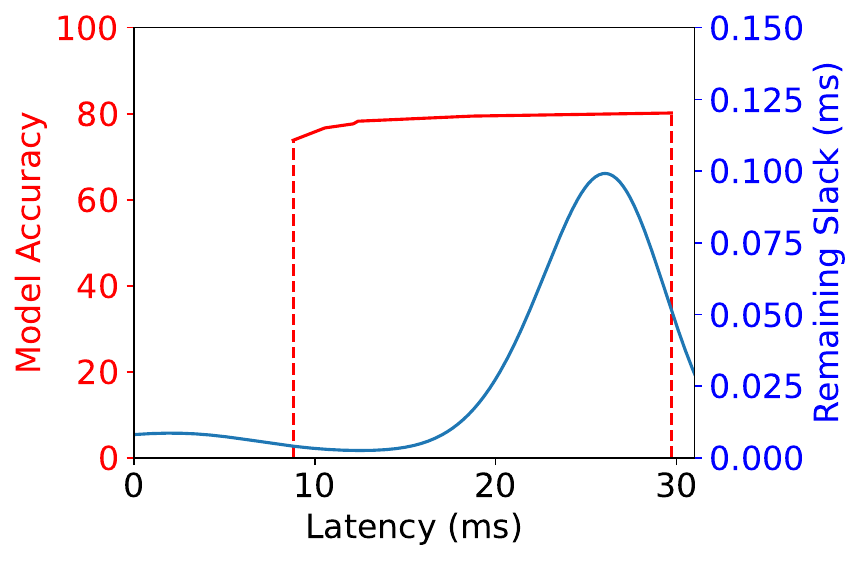}
        \caption{\small Clipper$^{+}$($78.25$)}
         \label{fig:appendix:clipper_78}
         \vspace{-0.01in}
    \end{subfigure}%
    \hfill
    \begin{subfigure}[b]{0.33\textwidth}
        \includegraphics[width=1\textwidth,keepaspectratio]{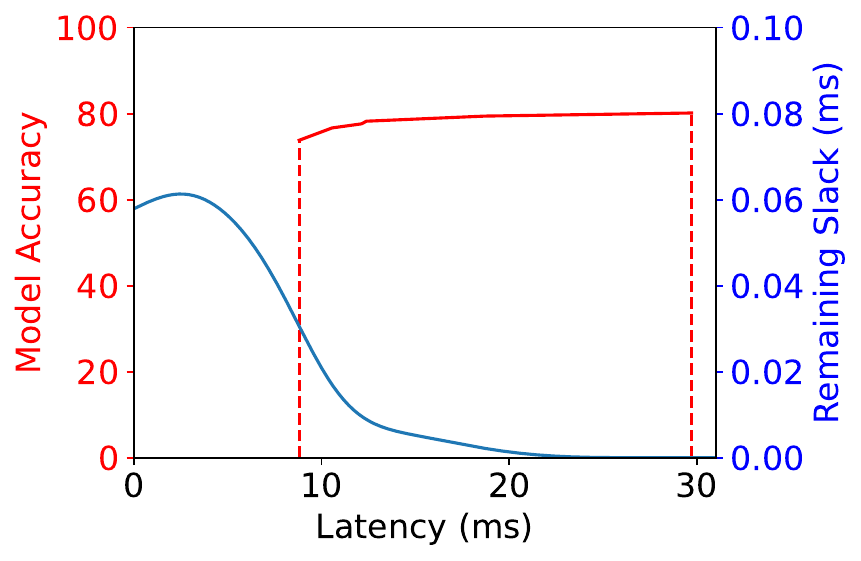}
        \caption{\small Clipper$^{+}$($79.44$)}
         \label{fig:appendix:clipper_79}
         \vspace{-0.01in}
    \end{subfigure}%
    \hfill
    \begin{subfigure}[b]{0.33\textwidth}
        \includegraphics[width=1\textwidth,keepaspectratio]{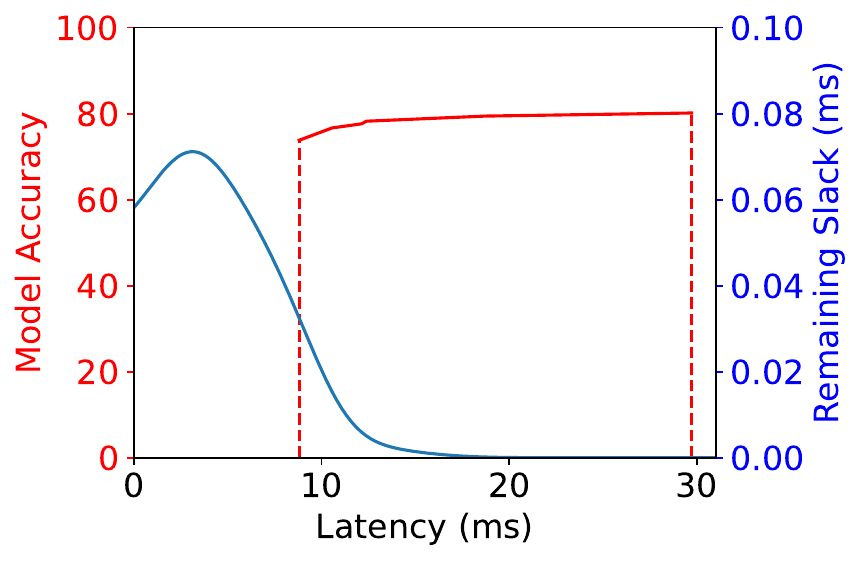}
        \caption{\small Clipper$^{+}$($80.16$)}
         \label{fig:appendix:clipper_80}
         \vspace{-0.01in}
    \end{subfigure}
    
    \caption{\small 
        \textbf{Serving fixed accuracy points}
    }
    \label{fig:appendix:clipper_plus_serving}
\end{figure*}

 For the $6$ Clipper$+$ experiments, the scheduler is constrained to provision all request-batches with just a single model choice. This greatly affects the system throughput, queueing delays and the SLO attainment. 
\begin{itemize}
    \item While the lower accuracy models ($73.82$, $76.69$ and $77.64$) have low inference latencies, they are able to serve the request batches within the SLO deadlines with low queueing delays. Hence they achieve $99\%+$ latency $SLO$ attainment. This can also be seen from Fig \ref{fig:appendix:clipper_plus_serving} (a-c) where the remaining slack latency is right-skewed and around $30ms$, which is the per-query latency deadline.
    \item On the other hand, higher accuracy models ($78.25$, $79.44$ and $80.16$) have significantly higher inference latencies. This leads to higher per-request serving time, higher queueing delays and lower SLO attainment. This can be seen from Fig \ref{fig:appendix:clipper_plus_serving} (d-f) where the distribution begins to shift leftwards towards lower remaining slack indicating significant queueing delays and lower SLO attainment.
\end{itemize}
The above results show that fixed-accuracy scheduling is insufficient to provide the best accuracy-SLO attainment tradeoff across different model choices, for any given arrival-trace.
\par
\begin{figure*}[!h]
	\centering
	\begin{subfigure}[b]{0.7\columnwidth} %
        \includegraphics[width=\columnwidth, height=0.6\textwidth]{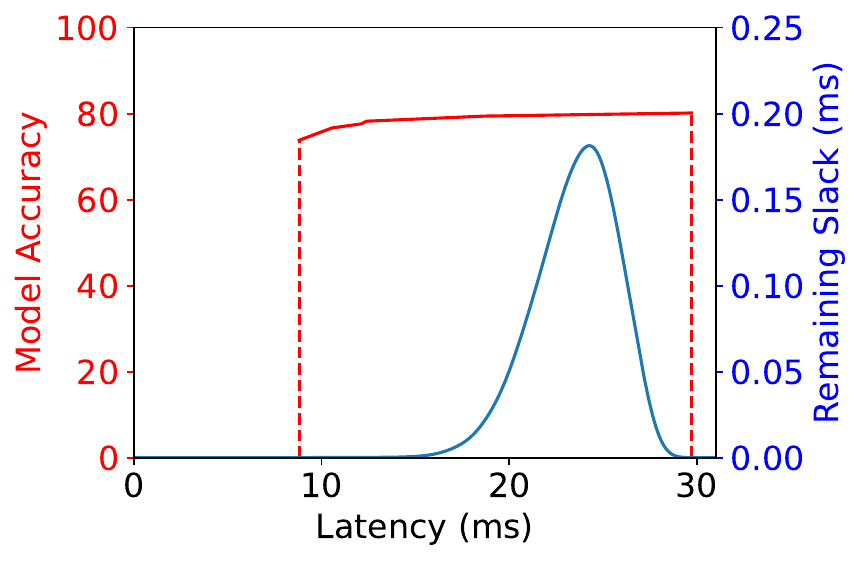}
	    \caption{ \syspol }
	    \label{fig:appendix:bucket}
	\end{subfigure}
    \caption{\small 
        \textbf{Serving multiple accuracy points}  
    }
	 \vspace{-0.2in}
\end{figure*}

 \syspol on the other hand performs dynamimc model selection throughout the models' latency range. This provides the scheduler several choices to maximize accuracy (higher accuracy models) during low loads and also leverage the lower accuracy models during high loads. This policy serves multiple accuracy points to help absorb queue build-up and mitigate queueing delays. Since Fig \ref{fig:appendix:bucket} shows that the density function is always below the supernetwork latency range, $>99\%$ queries meet their deadline, while also maximizing serving accuracy.

\end{document}